\documentclass[12pt]{iopart}

\usepackage[mathscr]{eucal} 
\usepackage{graphicx}
\usepackage{tikz} 
\usepackage{array}
\usepackage{float}
\usepackage[utf8]{inputenc}
\usepackage[colorlinks=true, urlcolor=blue,linkcolor=blue,citecolor=blue]{hyperref}
\usepackage{etoolbox}

\expandafter\let\csname equation*\endcsname\relax
\expandafter\let\csname endequation*\endcsname\relax
\usepackage{amsmath,amsfonts,amssymb,amsthm,amscd}

\makeatletter
\def\@mkboth#1#2{}
\newlength\appendixwidth
\preto\appendix{\addtocontents{toc}{\protect\patchl@section}}
\newcommand{\patchl@section}{%
  \settowidth{\appendixwidth}{\textbf{Appendix }}%
  \addtolength{\appendixwidth}{1.5em}%
  \patchcmd{\l@section}{1.5em}{\appendixwidth}{}{\ddt}%
}
\makeatother

\numberwithin{equation}{section}

\newcommand{\Z}{{\mathbb Z}}

\begin{document}

\title{Current correlations, Drude weights and large deviations in a box-ball system}

\author{Atsuo Kuniba$^1$, Grégoire Misguich$^{2}$ and Vincent Pasquier$^2$}

\address{$^1$ Institute of Physics,
  University of Tokyo, Komaba, Tokyo 153-8902, Japan}
\address{$^2$ Université Paris-Saclay, CNRS, CEA, Institut de physique théorique, 91191, Gif-sur-Yvette, France}
\eads{\mailto{atsuo.s.kuniba@gmail.com}, \mailto{gregoire.misguich@ipht.fr}
  \mailto{vincent.pasquier@ipht.fr}}

\begin{abstract}
  We explore several aspects of the current fluctuations and correlations in the box-ball system (BBS), an integrable cellular automaton in one space dimension. The state we consider is an ensemble of microscopic configurations where the box occupancies are independent random variables (i.i.d. state), with a given mean ball density. We compute several quantities exactly in such homogeneous stationary state: the mean value and the variance of the number of balls $N_t$ crossing the origin during time $t$, and the scaled cumulants generating function associated to $N_t$. We also compute two spatially integrated current-current correlations. The first one, involving  the long-time limit of the current-current correlations, is the so-called Drude weight and is obtained with thermodynamic Bethe Ansatz (TBA). The second one, involving  equal time current-current correlations is calculated using a transfer matrix approach. A family of generalized currents, associated to the conserved charges and to the different time evolutions of the models are constructed. The long-time limits of their correlations generalize the Drude weight and the second cumulant of $N_t$ and are found to obey nontrivial symmetry relations. They are computed using TBA and the results are found to be in good agreement with microscopic simulations of the model. TBA is also used to compute explicitly the whole family of flux Jacobian matrices. Finally, some of these results are extended to a (non-i.i.d.) two-temperatures generalized Gibbs state (with one parameter coupled to the total number of balls, and another one coupled to the total number of solitons).
\end{abstract}

\date{\today}

\tableofcontents

\section{Introduction}
\begin{figure}[h]
  \begin{center}
    \includegraphics[width=0.5\linewidth]{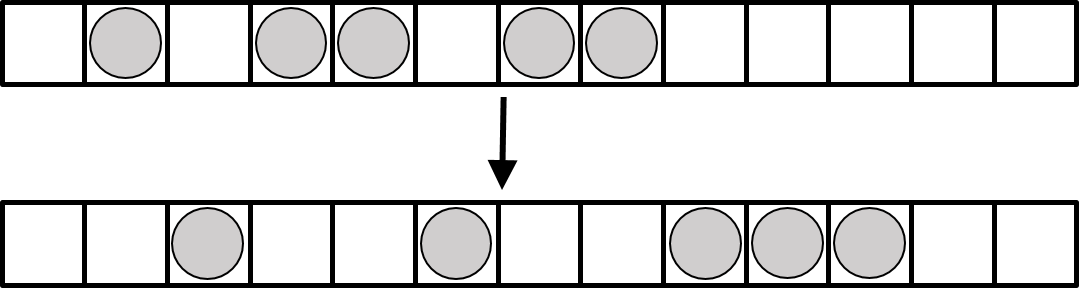}
    \caption{Top line: ball configuration. Bottom line: configuration at the next time step. We assumed here a carrier capacity $l\geq 3$.}
    \label{fig:dyn}
  \end{center}
\end{figure}

The box ball system (BBS) is an integrable cellular automaton introduced in 1990 by Takahashi and Satsuma \cite{TS90}. In this model some  ``balls'' occupy the sites (``boxes'') of a one-dimensional lattice  and propagate according to some simple deterministic rules, as explained below.
We start from an initial configuration of balls divided into $L$ boxes, with at most one ball per box. The configuration at the next time step is obtained by letting a ``carrier" travel through the system from left to right. Doing so, each time the carrier passes over an occupied box and if it has not reached its maximum load $l$, it loads the ball and leaves the box empty. Each time the carrier carries at least one ball and passes over an empty box it unloads a ball in the box.
An example is given in Fig.~\ref{fig:dyn}.\footnote{The dynamics can also be defined on a periodic system by initializing the carrier load to some suitable value (Proposition 5.1 of \cite{IKT12}).}

Despite its apparent simplicity this model possesses stable solitons with nontrivial scattering under collisions. It has an infinity of conserved quantities, and its very rich mathematical and integrable structures have attracted a lot of interest.
For example, the above combinatorial rule for the time evolution has its origin in the quantum $R$ matrix at $q=0$ as explained in \ref{app:eta_def}  (see \cite{IKT12} for a review).

An interesting family of problems arises when considering a statistical ensemble of random microscopic ball configurations \cite{LLP17,CKST18,KLO18,FG18,FNRW,LLPS19, CS19,CS20, KL20}.
In the simplest case the occupancies of the boxes can be taken to be independent and identically distributed (i.i.d.) random variables, and parameterized with a single parameter (the mean ball density).
Such a statistical ensemble corresponds to an homogeneous and stationary state, and many quantities, like the densities of the various types of solitons and their associated mean velocities, can be computed exactly using the thermodynamic Bethe Ansatz (TBA) \cite{KLO18,KL20,KMP20}.

It has also been shown that a hydrodynamic approach can accurately capture the long-time and large distance evolution of some inhomogeneous states. Due to the extensive number of conserved quantities in the model, the appropriate framework is the so-called generalized hydrodynamics (GHD).
Introduced a few years ago \cite{Castro-Alvaredo2016,Bertini2016,D19}, the GHD
is based on the assumption that the system is locally in an homogeneous state
with maximum entropy (generalized Gibbs ensemble (GGE) \cite{rigol_relaxation_2007}) and the approach
amounts to constructing and solving the set of continuity equations involving the local currents and local densities associated to the conserved quantities of an integrable system. It has been applied successfully to many integrable quantum and classical systems. On the classical side, we can mention for instance
the application of GHD to the hard-rods model \cite{DS17,DYC18}, to the Toda model \cite{doyon_generalized_2019}, to the sinh-Gordon model \cite{bastianello_generalized_2018}, to the BBS \cite{CS20,KMP20} and to the (higher rank) complete BBS \cite{kuniba_generalized_2021}.

In the context of BBS, the evolution triggered by an initial
domain-wall state with two different ball densities in the left half and the right half has been studied in details using GHD \cite{KMP20,kuniba_generalized_2021}.
In such a setup the BBS develops a series of plateaux in the variable $\zeta=x/t$ (space coordinate divided by time). The ball density and soliton content inside each plateau
as well as the location $\zeta_0$, $\zeta_1$, $\cdots$ of the steps between consecutive plateaux could be determined exactly using GHD.
In this domain-wall problem it has also been possible to go beyond the simplest hydrodynamic description by investigating some fluctuations effects. Due to the fact that the velocity of a given soliton is affected by the fluctuations of the densities of the other species of soliton, these velocities fluctuate and the step between two consecutive plateaux
is not perfectly sharp but broadened. This broadening has a diffusive scaling
and the width of the step number $k$ behaves as $\delta x \simeq t^{1/2}\Sigma_k$ with constants $\Sigma_k$ that have been calculated analytically. All these results for the domain wall problem have been also checked accurately using numerical simulations.

In the present work we are interested in the probability distribution of the number $N_t$ of balls
passing through the origin during a time $t$. We may take an initial domain wall state where
the left half is an i.i.d. state with ball density $p_{\rm left}>0$, and the right half initially empty ($p_{\rm right}=0$).
In such a case the number of transferred balls is also equal the number of balls in the right half at time $t$. But the ball propagation takes place only in the right direction
and there is no ``blocking effect" from existing balls in the BBS dynamics. So, $N_t$ does in fact not depend on the initial state in the right half of the system. The probability distribution of $N_t$ is therefore the same i) for the domain wall setup with ($p_{\rm left}=p$ and $p_{\rm right}=0$) and ii) in a uniform i.i.d. state with ball density $p_{\rm left}=p_{\rm right}=p$.

In the first step (Sec. \ref{ssec:c2}) we are interested in the second cumulant $c_2$ of $N_t$.
The result is obtained by two different methods, by a direct calculation and with
the TBA.
Next (Sec. \ref{ssec:D}) we  compute the Drude weight, which is defined here
in terms of the long-time limit of a spatially integrated current-current correlation.
This correlation is formally similar to the one defining the second cumulant of $N_t$
and the Drude weight is obtained using a TBA calculation which parallels that of the second cumulant.

A particularity of the model is to have a commuting family of temporal evolutions which are in duality with the conserved quantities.\footnote{Although the commuting time evolutions should in principle be a common feature of integrable systems, their implementation can be quite involved in general. This is one reason why the  BBS is a precious model, since it allows concrete descriptions
of the different time evolution, as well as their simulations.}
In Sec.~\ref{ssec:generalized_currents} the currents associated with  the conserved charges under the different temporal evolutions are introduced.
These depend on one index associated to a conserved charge, and another one associated to a time evolution. The generalized current operators turn out to be symmetric under the exchange of the two indices, highlighting this duality. 
We compute the long-time limit of the correlations associated to these generalized currents. They generalize $c_2$ and $D$,  and are shown to enjoy nontrivial symmetry relations among them.
In Sec.~\ref{ssec:drude_numerics} we report on numerical calculations of the Drude weights and the generalized current correlations using microscopic
simulations of the BBS, and a good agreement is found with the TBA results.
In Sec.~\ref{ssec:flux_jacobian} we compute the so-called flux Jacobian, which is a matrix playing an important role in generalized hydrodynamics.
The Sec.~\ref{ssec:f} discusses another quantity
defined in terms of spatially integrated current-current correlation, the variance of the total current.
Contrary to the Drude weight the correlations are now taken at equal time, and 
this variance is obtained using a transfer matrix approach.

In the second part of the paper (Sec. \ref{sec:ldf}) we compute the
scaled cumulants generating function
(SCGF) associated to $N_t$. It characterizes all the cumulants of the probability distribution \cite{touchette_large_2009}.
The Legendre transform of the SCGF, called large deviation rate function,
describes not only the typical fluctuations around its mean value, but also the rare events. The large deviation rate function
is compared with the results of numerical simulations. Finally, several results (second cumulant, Drude weight and SCGF)
are generalized to a more complex GGE, with two temperatures (\ref{sec:2TGGE}), one coupled to the total number of balls and the other one coupled to the total number of solitons.

\section{Current correlations and Drude weights}
\setcounter{footnote}{0}

We consider an i.i.d. homogeneous stationary state (characterized by some ball density $p=z/(z+1)<1/2$) and we study
the probability distribution of the number $N_t$ of balls crossing the origin between time 0 and time $t$, in the long time limit.
The capacity of the carrier which induces the dynamics is denoted by $l$.

\subsection{Mean current}
The mean value of $N_t$ grows linearly with time at a rate given by the mean ball current:
\begin{equation}
  \left< N_t \right> \sim t \; j^{(l)}.
\end{equation}
The mean ball current $j^{(l)}$
is a simple function of the soliton currents $j_k^{(l)}$
\begin{equation}
  j^{(l)}(z)= \sum_{k=1}^\infty k \;j_k^{(l)}.
  \label{eq:jkj}
\end{equation}
The factor $k$ above  reflects the fact that each type-$k$ solitons carries $k$ balls.
The mean soliton currents are products of the soliton velocities $v_k^{(l)}$ by the soliton densities $\rho_k$:
\begin{equation}
  j_k^{(l)}= \rho_k v_k^{(l)}.
\end{equation}
In integrable systems such densities can be obtained using the TBA \cite{yang_thermodynamics_1969}.
This approach, applied to the BBS \cite{KLO18,KMP20}, leads to a simple
expression for the mean soliton densities in terms of the ball fugacity $z=p/(1-p)$:
\begin{equation}
  \rho_k = \frac{z^k(1-z)^3(1+z^{k+1})}
  {(1+z)(1-z^k)(1-z^{k+1})(1-z^{k+2})}.
\end{equation}
As for the velocities, they obey a set of equations which reflect the fact
that the mean velocity of each soliton species is affected by the collisions with the solitons of the other species. The essential ideas were proposed in \cite{Zakharov_1971,El_2003,el_kinetic_2005},
to describe
soliton gases in the context of the Korteweg-de Vries and nonlinear Schrödinger equations.
In the case of BBS the equation for the effective velocities reads \cite{ferrari_soliton_2021} (see also \cite{CS20,KMP20})
\begin{equation}
  v^{(l)}_i = \kappa^{(l)}_i
  + \sum_{k=1}^\infty M_{i,k}(v^{(l)}_i-v^{(l)}_k)\rho_k\qquad (\forall i \ge 1),
  \label{eq:velocities_eq}
\end{equation}
where the bare soliton velocities are
\begin{equation}
  \kappa^{(l)}_i=\min(i,l)
  \label{eq:kappa}
\end{equation}
and the matrix $M$ (with indices from $[1,\infty]\times[1,\infty]$) encoding the shifts experienced by solitons during collisions is
\begin{equation}
  M_{i,k}=2\min(i,k).
  \label{eq:M}
\end{equation}
For i.i.d. states these equations could be solved explicitly \cite{KMP20} and the following expression
for the velocities were obtained:
\begin{equation}
  v^{(l)}_k = \frac{1+z^{l+1}}{1-z^{l+1}}v_{\min(k,l)},
  \qquad
  v_k = \frac{1+z}{1-z}k - \frac{2z(1+z)(1-z^k)}{(1-z)^2(1+z^{k+1})}.
  \label{eq:veff}
\end{equation}
Remark: $v_k^{(\infty)}=v_k$.
Combining the results above the mean ball current finally reduces to
\begin{equation}
  j^{(l)}(z)=\frac {z}{1-z}-\left( l+1 \right){\frac {{z}^{l+1}  }{1-{z}^{l+1}}}.
  \label{eq:jb}
\end{equation}

\subsection{Second cumulant}
\label{ssec:c2}
Next we are interested in the second cumulant of $N_t$, which turns out to be linear in $t$:
\begin{equation}
  \left<  \delta N_t^2 \right>= \left< N_t^2 \right>-\left< N_t \right>^2 \sim t \,c_{2}.
\end{equation}
This cumulant can be expressed as a time-integrated current-current correlation:
\begin{align}
  N_t&=\int_0^t j(0,s) ds,\\
  \left<  \delta N_t^2 \right>&=\int_0^t \int_0^t ds ds' \left<  j(0,s) j(0,s')\right>^c \sim t \,c_2,
\end{align}
where $j(x,s)$ is the current flowing from the site $x-1$ to the site $x$ at time $s$.
The superscript $(l)$ specifying the dynamics $T_l$ will often be  omitted  in what follows.
A continuous time notation is used here for clarity, but the BBS is a discrete time model and $\int_0^t ds$ is
in fact equivalent to $\sum_{s=0}^t$.
In the carrier picture for BBS, $j(x,s)$ is nothing but the load of the carrier on this link.
A general method to obtain such cumulants in integrable systems is described in 
\cite{myers_transport_2020,doyon_fluctuations_2020}.
It applies when all the cumulants scale linearly in time.
In what follows we apply this method to the case of the BBS, where it simplifies considerably.
\subsubsection{Correlations and sum rules.}

We start from the general sum rule (\ref{eq:sum_rule}) derived in \ref{sec:sr}
and specialize it to $f(x)=|x|$:
\begin{align}
  &\sum_{x=-\infty}^\infty |x|\left<\left( n(x,t)-n(x,0)\right)\left(n(0,t)-n(0,0)\right)\right>^c \nonumber\\
  &=-2\int_0^t \int_0^t ds ds' \left<  j(0,s) j(0,s')\right>^c.
\end{align}
Using the translation invariance (in space and time) of the correlators
appearing in the l.h.s we get
\begin{align}
  \left<  \delta N_t^2 \right>=&\frac{1}{2}\sum_{x=-\infty}^\infty |x|\left(
  \left< n(x,t)n(0,0)\right>^c
  +\left< n(-x,t)n(0,0)\right>^c
  -2\left< n(x,0)n(0,0)\right>^c
  \right) \nonumber \\
  &=\sum_{x=-\infty}^\infty |x|\left(
  \left< n(x,t)n(0,0)\right>^c
  -\left< n(x,0)n(0,0)\right>^c
  \right).
\end{align}
The equation above is essentially equivalent to (2.23) of \cite{mendl_current_2015} (see also (A.33) of \cite{doyon_fluctuations_2020}).
The second cumulant is then expressed as
\begin{equation}
  c_2 = \lim_{t\to \infty} \frac{1}{t}\sum_{x=-\infty}^\infty |x|\left(
  \left< n(x,t)n(0,0)\right>^c
  -\left< n(x,0)n(0,0)\right>^c
  \right).
  \label{eq:c2nn}
\end{equation}
The connected equal time correlation $ \left< n(x,0) n(0,0)\right>^c$ vanishes for $x\ne0$ in the i.i.d. state, so that
\begin{equation}
  c_2=\lim_{t\to\infty}\frac{1}{t}\sum_{x=-\infty}^\infty |x|\langle n(x,t) n(0,0)\rangle^c.
\end{equation}
Since the propagation only takes place in the right direction in the BBS, causality implies that the connected correlation $\langle n(x,t)n(0,0)\rangle^c$ vanishes
if $x<0$ (for $t>0$). So we have:
\begin{equation}
  \sum_{x=-\infty}^\infty |x|\langle n(x,t) n(0,0)\rangle^c
  =\sum_{x=-\infty}^\infty x\langle n(x,t)n(0,0)\rangle^c.
\end{equation}
We differentiate with respect to time, and use the local conservation law $\frac{d}{dt} n(x,t)= j(x,t)- j(x+1,t)$:
\begin{align}
  \frac{d}{dt}\sum_{x=-\infty}^\infty |x|\langle n(x,t) n(0,0)\rangle^c
  &=\sum_{x=-\infty}^\infty x\langle \left( j(x,t) -j(x+1,t)\right) n(0,0)\rangle^c \label{eq:c2A}\\
  &=\sum_{x=-\infty}^\infty \langle j(x,t) n(0,0)\rangle^c 
  =\sum_{x=-\infty}^\infty \langle j(0,0) n(x,-t)\rangle^c.\label{eq:c2B}
\end{align}
We recognize the conserved  charge $Q=\sum_{x=-\infty}^\infty n(x,-t)$ in the r.h.s., which means that (\ref{eq:c2A})-(\ref{eq:c2B}) is independent of time. It can be evaluated at $t=0$, or, instead, averaged over time. This yields two equivalent formulations of $c_2$ in terms of integrated current-density correlations:
\begin{align}
  c_2&=\sum_{x=-\infty}^\infty \langle j(x,0) n(0,0)\rangle^c \label{eq:c2_jn0} \\
  &=\lim_{t\to\infty}\frac{1}{t}\int_0^t ds\sum_{x=-\infty}^\infty \langle j(x,s) n(0,0)\rangle^c \label{eq:c2_jnt}.
\end{align}

\subsubsection{Direct calculation.}
In a GGE where $\beta$ is the inverse temperature associated to the total number of balls (and $z=\exp(-\beta)$),
(\ref{eq:c2_jn0}) can be written as
\begin{equation}
  c_2=-\frac{\partial j}{\partial \beta}
  =z\frac{\partial j}{\partial z}.
\end{equation}
Using (\ref{eq:jb}) we directly obtain
\begin{equation}
  c_2=\frac{z}{(1-z)^2}-(l+1)^2\frac{z^{l+1}}{(1-z^{l+1})^2}.
  \label{eq:c2b}
\end{equation}
The above formula has been compared with numerical simulation of the BBS. The mean value and the second cumulant turn out to be in good agreement with the theoretical values (Tab.~\ref{tab:c2}).
A generalization of this result to a two-temperature GGE
is given in (\ref{eq:c2l}).

\begin{table}
  \center
  \begin{tabular}{|c|c|c|c|c|c|}
    \hline
    $p$   & $l$  & $t$    & $c_2$  (\ref{eq:c2b}) & $\left< \delta N_t^2 \right>/t$ numerics \\
    \hline
    0.2 & 4  & 6000 & 0.41998               & 0.419                                      \\
    0.4 & 4  & 2000 & 1.63352               & 1.63                                       \\
    0.3 & 10 & 2000 & 1.30166               & 1.29                                       \\
    \hline
  \end{tabular}
  \caption{Variance of the number of transferred balls for different values of the density $p$ and capacity $l$. Comparison between (\ref{eq:c2b}) and numerical results (last column).}
  \label{tab:c2}
\end{table}

\subsubsection{Second cumulant using TBA.}

We will now compute $c_2$ by a different approach. The total current $J$ and the total number
of balls $Q$ can be decomposed into a mean value plus a fluctuating part: $J=Lj+\delta J$, $Q=Lp+\delta Q$.
With these definitions (\ref{eq:c2_jnt}) gives
\begin{equation}
  c_2=L^{-1}\lim_{t\to\infty}\frac{1}{t}\int_0^t ds\left< \delta J(s) \delta Q(0)\right>.
  \label{eq:c2_JQ}
\end{equation}
The pseudoenergy
$\epsilon_i=-\ln\left(\frac{2E_i-E_{i+1}-E_{i-1}}{L-2E_i}\right)$
can be defined for each microscopic configuration, from the energies $E_i$, $E_{i-1}$ and $E_{i+1}$.
See \ref{app:eta_def} or [eq.~(2.7), \cite{KMP20}]
for the explanation of $E_i$.\footnote{
The pseudoenergies can be used to write down the (fermionic) free energy
$F=-\sum_k \ln(1+e^{-\epsilon_k})$ [(3.15) in \cite{KMP20}] and the mode occupancies
are $n_k=dF/d(\epsilon_k)=(1+e^{\epsilon_k})^{-1} =(1+1/y_k)^{-1}$ with $y_k=e^{-\epsilon_k}$.
}
These energies are conserved under the time evolution, and so are the pseudoenergies.
The pseudoenergies can however fluctuate from configuration to configuration in a GGE.
If we define $\delta\epsilon_i = L^{1/2}(\epsilon_i -\bar\epsilon_i )$,
with $\bar\epsilon_i=\left<\epsilon_i\right>$, these fluctuations 
have diagonal correlations \cite{KMP20}
\begin{equation}
  \left< \delta \epsilon_i \delta \epsilon_j\right>=\delta_{ij}(1+e^{\bar\epsilon_i})/\sigma_i,
  \label{eq:ee}
\end{equation}
where
\begin{equation}
\sigma_i=1-\sum_k M_{ik}\rho_k
\label{eq:sigma_rho}
\end{equation}
is the hole density
and $M$ was given in (\ref{eq:M}).\footnote{The values of $\sigma_i$ in the i.i.d. state are given by
(\ref{roaz}) with $a=z$ (or (3.25) in \cite{KMP20}).}
Note also that the expectation values of the pseudoenergies
are given by $\bar\epsilon_i=-\ln \left(\rho_i / \sigma_i\right)$.
The relation (\ref{eq:ee}) was checked numerically (see Tab.~\ref{tab:epsilon_correl}).
\begin{table}[H]
  \begin{center}\begin{tabular}{|c|c|c|c|}
      \hline
      i & j& $\delta_{ij}(1+e^{\bar\epsilon_i})/\sigma_i$ & $\left< \delta \epsilon_i \delta \epsilon_j\right>$ numerics\\
      \hline
      1 & 1 & 8.01282 & 8.0117\\
      1 & 2 & 0       & 0.00027\\
      1 & 3 & 0       & -0.00079\\
      2 & 2 & 27.2183 & 27.223\\
      2 & 3 & 0       & -0.0053\\
      3 & 3 & 65.5367 & 65.588 \\
      4 & 4 & 131.789 & 132.07 \\
      5 & 5 & 238.454 & 239.37 \\
      \hline
    \end{tabular}
  \end{center}\caption{
    Pseudoenergy correlations for ball density $p=0.4$: comparison between
    the TBA result (r.h.s of (\ref{eq:ee})), and numerical simulations (system size $L=8\,10^4$ and average over $N_{\rm samples}=10^8$ configurations).
The numerical results correspond to
$L \left<\ln\left(\frac{2E_i-E_{i+1}-E_{i-1}}{L-2E_i}\right)
\ln\left(\frac{2E_j-E_{j+1}-E_{j-1}}{L-2E_j}\right)\right>^c$. As for the expectation values
$\langle \epsilon_i\rangle$, the simulation results agree with $-\ln \left(\rho_i / \sigma_i\right)$ with very good accuracy (data not shown).
}\label{tab:epsilon_correl}
\end{table}

In turn,  the fluctuations of the other quantities can be related
to the fluctuations $\delta \epsilon_j$.
In a large system $\delta J/L$ and
$\delta Q/L$ are typically small, of order $\mathcal{O}(L^{-1/2})$, and we may linearize the relation between the ball density fluctuation and the pseudoenergies, 
\begin{equation}
  \delta Q /\sqrt{L} = \sum_{i=1}^\infty \delta \epsilon_i \frac{\partial \rho_b}{\partial \bar\epsilon_i}, \label{eq:Q}
\end{equation}
as well as the relation between the ball current density fluctuation and the pseudo energies
\begin{equation}
  \delta J / \sqrt{L} = \sum_{i=1}^\infty \delta \epsilon_i \frac{\partial j_b}{\partial \bar\epsilon_i} +\cdots. \label{eq:J}
\end{equation}
In the equations above the ball density was denoted by $\rho_b$ and the ball current by $j_b$.
In (\ref{eq:J}) we have decomposed a current fluctuation, which is {\em not} conserved in time, in terms of fluctuations $\delta \epsilon_i$ of the pseudoenergies. Since each $\epsilon_i$ is a function of the $\{E_k\}$, the pseudoenergies and
their fluctuations $\delta \epsilon_i$ are configuration-dependent but independent of time. So, by decomposing $\delta J$ over the $\delta \epsilon_i$ we have dropped all the time dependence in $\delta J$ and only the conserved component of the current has been kept (hence the dots in (\ref{eq:J})). Following the idea of hydrodynamics projection onto the space of conserved quantities \cite{D19} we are here restricting ourselves to the part of the current which is constant in time and can be expressed in terms of the conserved energies.
It is of course legitimate to do so in order to compute $c_2$, thanks to the long-time limit in (\ref{eq:c2_JQ}). Note that, in contrast,
nothing was dropped in (\ref{eq:Q}) since the total number of balls  (and thus also $\delta Q$) is a conserved quantity.

Replacing (\ref{eq:Q}), (\ref{eq:J}) and (\ref{eq:ee}) in (\ref{eq:c2_JQ}) yields 
\begin{equation}
  c_2 = \sum_{i=1}^\infty \frac{\partial \rho_b}{\partial \bar\epsilon_i}\,
  \frac{\partial j_b}{\partial \bar\epsilon_i}\,
  \frac{1+e^{\bar\epsilon_i}}{\sigma_i}.
  \label{eq:c2tba}
\end{equation}
What remains to be done is to compute the derivatives $\frac{\partial \rho_b}{\partial \bar\epsilon_i}$
and $\frac{\partial j_b}{\partial \bar\epsilon_i}$.

Once the pseudoenergies associated to the GGE are known we can introduce the  ``dressing'' operation, which is standard in the framework of TBA. 
A set of quantities $o_i$ labelled by some index $i \in [1,\infty]$ representing a soliton size
can be grouped into a column vector $o=(o_1,o_2,\cdots)$. We then define the ``dressing matrix'' $G$ and construct a new dressed vector $o^{\rm dr}$ \cite{KMP20}:
\begin{align}
  o^{\rm dr}&=G o, \\
  G &= \left(1+M\hat y\right)^{-1},
  \label{eq:dress_def}
\end{align}
where $\hat y$ is the diagonal matrix with elements $y_j=\exp(-\bar\epsilon_j)$.\footnote{The diagonal entries of the matrix $M$ do not enter the velocity equation (\ref{eq:velocities_eq}) and there is therefore some freedom to redefine its diagonal part. This leads to some freedom in the definition of the dressing operation and the present choice (also used in \cite{KMP20}) is not the same as in \cite{D19}.
With (\ref{eq:dress_def}) we have
$o^{\rm dr}=o-M{\hat y}o^{\rm dr}$. In \cite{D19}
the dressing operation (denoted with a prime) is instead defined by
$o^{\rm dr'}=o+T{\hat n} o^{\rm dr'}$ where $T=1-M$
and $\hat n$ is the diagonal matrix with entries $n_i=\rho_i/(\sigma_i+\rho_i)$.
The two dressing definitions lead to the same effective soliton speeds $v_i^{(l)}=\kappa_i^{(l) {\rm dr}} / (1^{\rm dr})_i=\kappa_i^{(l) {\rm dr'}} / (1^{\rm dr'})_i$ and can both be used to obtain physical quantities.
}

The basic relations that will be used in the following calculation are
\begin{align}
y_i = e^{-\bar\epsilon_i} = \frac{\rho_i}{\sigma_i}, \qquad 
\sigma_i v^{(l)}_i = (G\kappa^{(l)})_i  \; \bigl(= (\kappa^{(l)\mathrm{dr}})_i\bigr).
\end{align}
From $G + GM\hat{y} = \mathrm{Id}$, we find that $GM$ is symmetric:
\begin{align}
GM = \hat{y}^{-1}-G\hat{y}^{-1} = \hat{y}^{-1}- (\hat{y}+ \hat{y}M\hat{y})^{-1} = (GM)^t.
\label{eq:GM_sym}
\end{align}
Its $(i,j)$ element is concretely expressed as
\begin{align}
(GM)_{ij} &= 2\sum_kG_{ik}\min(k,j) = 2(G\kappa^{(j)})_i = 2\sigma_i v^{(j)}_i,
\label{eq:GM}
\end{align}
where the invariance of the last expression under the interchange 
$i \leftrightarrow j$ can also be confirmed at the level of explicit formulae,
see (\ref{eq:j_hole_sigma}). 
Using $\frac{\partial y_n}{\partial \bar\epsilon_i} = -\delta_{in}y_i$, we get
\begin{align}\label{df1}
\frac{\partial G_{jk}}{\partial \bar\epsilon_i} 
&= -\sum_{m,n}G_{jm}\frac{\partial (M\hat{y})_{m n}}{\partial \bar\epsilon_i}G_{n k}
= \sum_m G_{jm}M_{mi}y_iG_{ik} 
\nonumber \\
&= (GM)_{ji}y_i G_{ik}
= (GM)_{ij}y_i G_{ik} = 2\sigma_i v^{(j)}_i y_i G_{ik}= 2\rho_i v^{(j)}_i G_{ik}.
\end{align}
Let us introduce $\eta^{(l)}_j$, which  includes, 
due to $v^{(l=1)}_k=1$, the density and the current of balls as special cases:
\begin{align}\label{etadef}
\eta^{(l)}_j = \sum_k \min(j,k)\rho_k v^{(l)}_k,
\qquad 
\eta^{(1)}_\infty = \rho_b, \qquad \eta^{(l)}_\infty = j_b.
\end{align}
It is expressed in terms of $G$ as 
\begin{align}
  \eta^{(l)}_j &
  = \sum_k \min(j,k) y_k \sigma_k v^{(l)}_k
  = \frac{1}{2}\sum_k (M\hat{y})_{jk} (G\kappa^{(l)})_k
  \nonumber \\
  &= \frac{1}{2}\sum_k(-\mathrm{Id}+ G^{-1})_{jk}(G\kappa^{(l)})_k
  = -\frac{1}{2}\sum_kG_{jk}\kappa^{(l)}_k + \frac{1}{2}\kappa^{(l)}_j = \eta^{(j)}_l,
  \end{align}
  where the last equality is due to the symmetry (\ref{eq:GM_sym}).
  The symmetry of $\eta$ under the exchange of the upper and lower indices will play an important role in Sec.~\ref{ssec:generalized_currents}. 
By means of (\ref{df1}),  the $\bar\epsilon_i$ derivative is calculated as
\begin{align}
\frac{\partial \eta^{(l)}_j}{\partial \bar\epsilon_i} 
= -\frac{1}{2}\sum_k\frac{\partial G_{jk}}{\partial \bar\epsilon_i}\kappa^{(l)}_k
= -\sum_k \rho_i v^{(j)}_i G_{ik}\kappa^{(l)}_k
=  -\rho_i \sigma_i v^{(j)}_iv^{(l)}_i. 
\end{align}
From this with $j= \infty$ and (\ref{etadef})  it follows that 
\begin{align}
\frac{\partial \rho_b}{\partial \bar\epsilon_i} &= -\rho_i\sigma_i v^{(\infty)}_i,
\label{eq:drbdei} \\
\frac{\partial j_b}{\partial \bar\epsilon_i} &= -\rho_i\sigma_i v^{(\infty)}_iv^{(l)}_i.\label{eq:djbdei}
\end{align}

The results (\ref{eq:drbdei}) and (\ref{eq:djbdei}) can be inserted into (\ref{eq:c2tba}):
\begin{equation}
  c_2 = \sum_{i \ge 1}\rho_i\sigma_i(\rho_i+\sigma_i)(v^{(\infty)}_i)^2v^{(l)}_i.
  \label{eq:c2tba_2}
\end{equation}
This expression can be checked to agree with (\ref{eq:c2b}).

\subsection{Drude weights} 
\label{ssec:D}
The Drude weight is an important quantity in the field of transport.
This linear response coefficient characterizes the increase of the current
in presence of a force which couples to the charges. The recent years have seen very important progress in the understanding of this quantity in one-dimensional integrable systems \cite{ilievski_microscopic_2017,ilievski_ballistic_2017,doyon_drude_2017,bulchandani_bethe-boltzmann_2018,krajnik_anisotropic_2021,de_nardis_correlation_2022}.
$D$ can be defined in terms of a time average current-current correlation function :\footnote{Using the sum rule (\ref{eq:sum_rule}) with the choice $f(x)=x^2$, (\ref{eq:Djj}) can also be written
\begin{equation}
  D=\lim_{t \to \infty} \frac{1}{t^2}\sum_x x^2\left< n(x,t) n(0,0)\right>^c,
\end{equation} which is another useful definition of $D$.}
\begin{equation}
  D=\lim_{t \to \infty} \frac{1}{t}\int_0^t ds \sum_x \left< j(x,s) j(0,0)\right>^c.
  \label{eq:Djj}
\end{equation}
Noting the close similarity with (\ref{eq:c2_jnt}), we repeat the hydrodynamic projection of the current fluctuations onto the pseudoenergy fluctuations, which lead to (\ref{eq:c2tba}). It yields to
\begin{equation}
  D=\sum_{i=1}^\infty \left(\frac{\partial j_b}{\partial \bar\epsilon_i}\right)^2\frac{1+e^{\bar\epsilon_i}}{\sigma_i}
  \label{eq:Dtba}
\end{equation}
and the derivatives $\frac{\partial j_b}{\partial \bar\epsilon_i}$ are again
given by (\ref{eq:djbdei}). We thus obtain a TBA expression for the Drude weight
\begin{equation}
D = \sum_{i \ge 1}\rho_i\sigma_i(\rho_i+\sigma_i)(v^{(\infty)}_iv^{(l)}_i)^2.
\label{dsum}
\end{equation}

The expression for the density correlation $R=\sum_x \left< n(x,t) n(0,0) \right>^c$  has a form  similar to that of $c_2$ and $D$.
Thanks to the conservation of the total number of balls $R$ is independent of time and in an i.i.d. state it is trivially equal to $p(1-p)$. Following the reasoning leading
to  (\ref{eq:c2tba_2}) and (\ref{dsum}) we get
\begin{equation}
  p(1-p)=\sum_{i \ge 1}\rho_i\sigma_i(\rho_i+\sigma_i)(v^{(\infty)}_i)^2.
  \label{eq:p1mp}
\end{equation}
For more details about the above formula and their expression in matrix forms we refer the readers to Sec.~\ref{sssec:cov_drude_mat}.
The structure of these formulae
is very similar to the results for $c_2$ and $D$ in the Lieb-Liniger model,
 obtained by Doyon and Spohn (eqs. (1.2) and (1.3) of \cite{doyon_drude_2017}).
 In the latter work a central idea is to implement the long-time limit in the definition of $D$ via the hydrodynamic projection
(see (2.18) of \cite{doyon_drude_2017}).
This approach amounts to projecting the observable of interest (for $D$, the current) via a suitable scalar product into the time-invariant subspace, spanned by the conserved charges. As mentioned in the paragraph before (\ref{eq:c2tba}), in the approach presented here the projection is implemented by decomposing the current fluctuation over the conserved variables $\epsilon_i$. The relation (\ref{eq:ee})
shows that the pseudoenergies form a convenient orthogonal basis in the subspace of conserved quantities.

\begin{figure}[H]
  \begin{center}
    \includegraphics[width=0.7\linewidth]{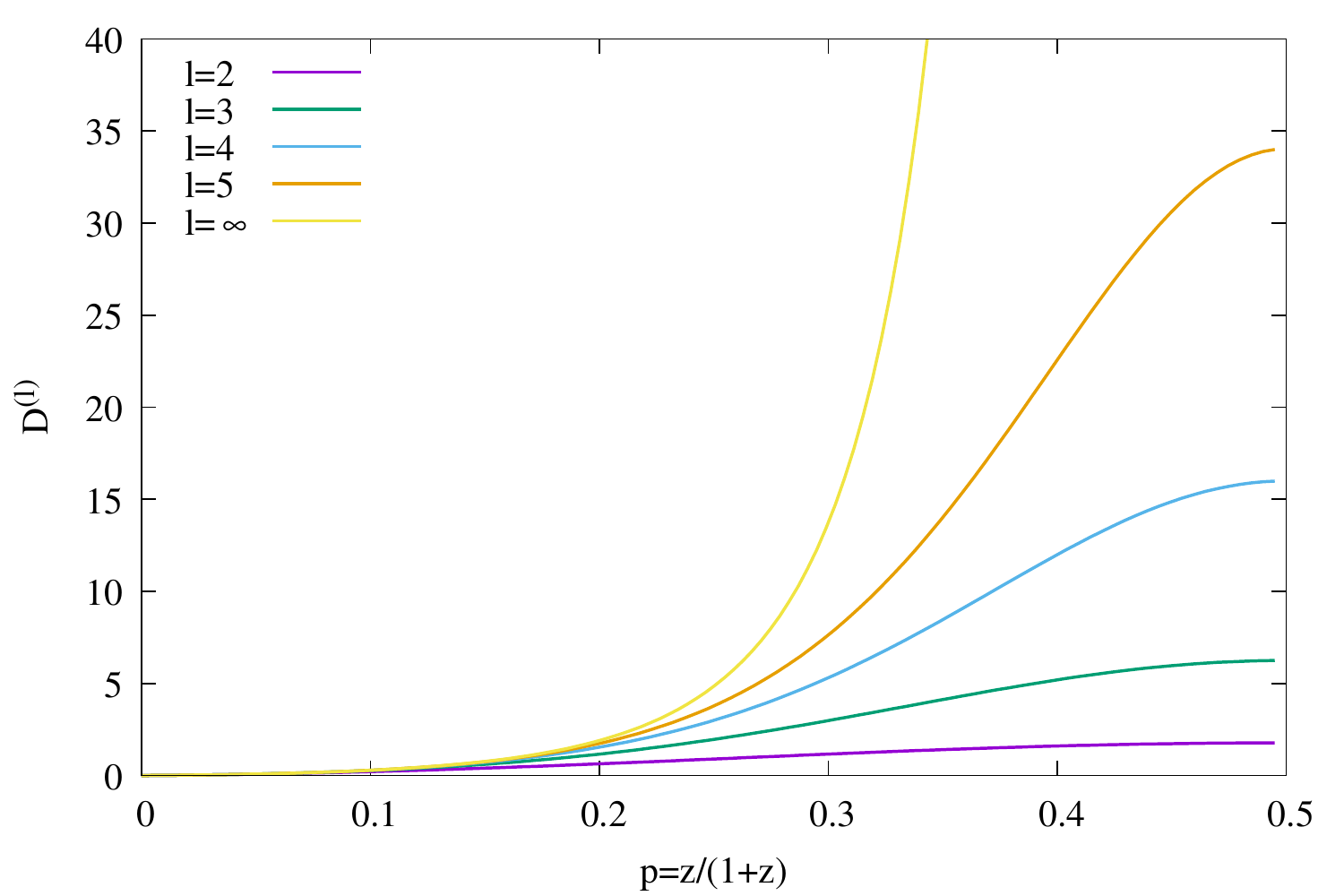}
    \caption{Drude weight $D^{(l)}$ for $l=2,3,4,5$ and $l=\infty$, plotted as a function of the ball density $p$. For $l\geq 2$ the Drude weight is linear in $p$ at low density: $D^{(l)}=p+11p^2+\mathcal{O}(p^3)$.
    }\label{fig:Dn}
  \end{center}
\end{figure}

Aided by some computer algebra system it was possible to
obtain explicit formulae for the Drude weight as a function of $z$ for a few values of the carrier capacity $l$, denoted by $D^{(l)}$ :
\begin{align}
D^{(2)} &= \frac{z(1-z)(1-z^2)^3(1+11z+11z^3+z^4)}{(1-z^3)^3(1-z^4)},
\\
D^{(3)} & = \frac{z(1-z)(1-z^2)^2(1-z^3)}{(1-z^4)^3(1-z^6)}
(1 + 11 z + 44 z^2 + 29 z^3 + 30 z^4 + 29 z^5 + 44 z^6 + 11 z^7 + z^8),
\\
D^{(4)} &= \frac{z(1-z^2)^4(1-z^3)}{(1-z^5)^3(1-z^6)(1-z^8)}
(1 + 10 z + 35 z^2 + 117 z^3 + 68 z^4 + 254 z^5 + 95 z^6 + 357 z^7 
\nonumber \\
& \qquad \qquad + 
  126 z^8 + 357 z^9 + 95 z^{10} + 254 z^{11} + 68 z^{12} + 117 z^{13} + 
  35 z^{14} + 10 z^{15} + z^{16}),
  \\
  D^{(5)} & = \frac{z(1-z)(1-z^2)^2(1-z^4)(1-z^5)}{(1-z^6)^3(1-z^8)(1-z^{10})}
  (1 + 11 z + 44 z^2 + 140 z^3 + 355 z^4 + 406 z^5 + 480 z^6 
  \nonumber \\
  &+ 
  443 z^7 + 633 z^8 + 714 z^9 + 896 z^{10} + 714 z^{11} + 633 z^{12} + 
  443 z^{13} + 480 z^{14} + 406 z^{15} 
  \nonumber \\
  &+ 355 z^{16} + 140 z^{17} + 44 z^{18} + 
  11 z^{19} + z^{20}).
\end{align}
The functions above as well as $D^{(\infty)}$ are plotted as a function  of the ball density
$p=z/(1+z)$ in Fig.~\ref{fig:Dn}. When the density is slightly below 0.5 we note a rapid increase of the Drude weight with $l$. Since $l$ acts as a cutoff on the effective speed of the solitons of size $k\geq l$, this indicates that the large solitons have a dominant contribution to the Drude weight.

Going further one can obtain an expression for $D^{(l)}$ for generic $l$ with finite number of terms. To this end we first set $W_i = \rho_i\sigma_i(\rho_i+\sigma_i)v^2_i$.
Then we have 
\begin{equation}\label{Asum}
\begin{split}
\sum_{i<l}W_i &= \frac{zA_l}{(1+z)^2(1-z^l)^2(1-z^{l+1})^2},
\\
A_l &=1+3z^{2l}+8 z^{2l+1}+3z^{2l+2}+z^{4l+2}
-(l+1)^2(z^{l+2}+z^{3l}) 
\\
&+(l^2-4)(z^{l+1}+z^{3l+1})
+(l-1)(l+3)(z^l+z^{3l+2}) - l^2(z^{l-1}+z^{3l+3}).
\end{split}
\end{equation}
Note that $A_0=A_1=0$ indeed holds.
It leads to 
\begin{align}
D^{(1)} = \sum_{i \ge 1}W_i(v^{(1)}_i)^2 = \sum_{i \ge 1} W_i 
= \frac{zA_\infty}{(1+z)^2} = \frac{z}{(1+z)^2}.
  \label{eq:D1}
\end{align}
By means of this, the infinite sum (\ref{dsum}) is reduced to a finite one as
\begin{align}
D^{(l)} &= \sum_{i<l}W_i(v^{(l)}_i)^2 + (v^{(l)}_l)^2\sum_{i \ge l}W_i
\\
& = \sum_{i<l}W_i(v^{(l)}_i)^2 + (v^{(l)}_l)^2\Bigl(\frac{z}{(1+z)^2}
-\sum_{i<l} W_i\Bigr),
\label{eq:Dl}
\end{align}
where the second term is known by (\ref{Asum}).
Or equivalently,  
\begin{align}
D^{(l)} = \left(\frac{1+z^{l+1}}{1-z^{l+1}}\right)^2
\left(
\sum_{i<l}W_i(v_i^2-v_l^2) + v_l^2 \frac{z}{(1+z)^2}
\right).
\end{align}
We note that in the half filled limit the Drude weight tends
to a simple value
\begin{align}
\lim_{z \rightarrow 1} D^{(l)} = \left(\frac{l(l+2)}{6}\right)^2.
\end{align}
This is consistent with the observation that
$D^{(\infty)}$ diverges when $z\to 1$ (see Fig.~\ref{fig:Dn}). 
To conclude this section we mention that some generalization of these results to a two-temperature GGE is discussed in \ref{sec:2TGGE}.

\subsection{Generalized current correlations and their symmetries}
\label{ssec:generalized_currents}

Thus far we have considered 
$c_2$ (\ref{eq:c2_JQ}), $D$ (\ref{eq:Djj}) and $R$ (\ref{eq:p1mp}).
They are all associated with the number balls, which is one special case $E_\infty$ of the 
conserved quantities $E_j$ with $j=1,2,\ldots$ \cite{KMP20}.
Here we discuss a natural generalization corresponding to the whole family $\{E_j\}$.

\begin{figure}
  \begin{center}
  \begin{picture}(350,85)(-55,-55)
    \put(0,12){0}\put(30,12){0}\put(60,12){0}\put(90,12){1}\put(120,12){1}\put(150,12){1}
    \put(180,12){0}\put(210,12){0}\put(240,12){1}\put(270,12){1}\put(300,12){0}
    \put(-15,-15){\put(-25,14){$T_3$}
    \put(0,12){2}\put(30,12){1}\put(60,12){0}\put(90,12){0}\put(120,12){1}\put(150,12){2}
    \put(180,12){3}\put(210,12){2}\put(240,12){1}\put(270,12){2}\put(300,12){3}\put(330,12){2}
    }
    \multiput(0,0)(0,-30){2}{\multiput(2,0)(30,0){11}{
    \put(-10,0){\line(1,0){20}}\put(0,9){\line(0,-1){18}}
    }}
    \put(0,-30){
    \put(0,12){1}\put(30,12){1}\put(60,12){0}\put(90,12){0}\put(120,12){0}\put(150,12){0}
    \put(180,12){1}\put(210,12){1}\put(240,12){0}\put(270,12){0}\put(300,12){1}
    }
    \put(-15,-30){\put(-35,12){\color{red}$\hat{\eta}^{(3)}_2(x)$}
    \put(0,12){\color{red}1}\put(30,12){\color{red}0}\put(60,12){\color{red}0}
    \put(90,12){\color{red}0}\put(120,12){\color{red}1}\put(150,12){\color{red}2}
    \put(180,12){\color{red}2}\put(210,12){\color{red}1}\put(240,12){\color{red}0}
    \put(270,12){\color{red}1}\put(300,12){\color{red}2}\put(330,12){\color{red}1}
    }
    \put(-15,-45){\put(-25,10){$T_2$}
    \put(0,12){1}\put(30,12){2}\put(60,12){2}\put(90,12){1}\put(120,12){0}\put(150,12){0}
    \put(180,12){0}\put(210,12){1}\put(240,12){2}\put(270,12){1}\put(300,12){0}\put(330,12){1}
    }
    \put(0,-60){
    \put(0,12){0}\put(30,12){1}\put(60,12){1}\put(90,12){1}\put(120,12){0}\put(150,12){0}
    \put(180,12){0}\put(210,12){0}\put(240,12){1}\put(270,12){1}\put(300,12){0}
    }
    \end{picture}
    \begin{picture}(350,85)(-55,-55)

      \put(0,12){0}\put(30,12){0}\put(60,12){0}\put(90,12){1}\put(120,12){1}\put(150,12){1}
      \put(180,12){0}\put(210,12){0}\put(240,12){1}\put(270,12){1}\put(300,12){0}
      \put(-15,-15){\put(-25,14){$T_2$}
      \put(0,12){1}\put(30,12){0}\put(60,12){0}\put(90,12){0}\put(120,12){1}\put(150,12){2}
      \put(180,12){2}\put(210,12){1}\put(240,12){0}\put(270,12){1}\put(300,12){2}\put(330,12){1}
      }
      \multiput(0,0)(0,-30){2}{\multiput(2,0)(30,0){11}{
      \put(-10,0){\line(1,0){20}}\put(0,9){\line(0,-1){18}}
      }}
      \put(0,-30){
      \put(0,12){1}\put(30,12){0}\put(60,12){0}\put(90,12){0}\put(120,12){0}\put(150,12){1}
      \put(180,12){1}\put(210,12){1}\put(240,12){0}\put(270,12){0}\put(300,12){1}
      }
      \put(-15,-30){\put(-35,12){\color{red}$\hat{\eta}^{(2)}_3(x)$}
      \put(0,12){\color{red}1}\put(30,12){\color{red}0}\put(60,12){\color{red}0}
      \put(90,12){\color{red}0}\put(120,12){\color{red}1}\put(150,12){\color{red}2}
      \put(180,12){\color{red}2}\put(210,12){\color{red}1}\put(240,12){\color{red}0}
      \put(270,12){\color{red}1}\put(300,12){\color{red}2}\put(330,12){\color{red}1}
      }
      \put(-15,-45){\put(-25,10){$T_3$}
      \put(0,12){2}\put(30,12){3}\put(60,12){2}\put(90,12){1}\put(120,12){0}\put(150,12){0}
      \put(180,12){1}\put(210,12){2}\put(240,12){3}\put(270,12){2}\put(300,12){1}\put(330,12){2}
      }
      \put(0,-60){
      \put(0,12){0}\put(30,12){1}\put(60,12){1}\put(90,12){1}\put(120,12){0}\put(150,12){0}
      \put(180,12){0}\put(210,12){0}\put(240,12){1}\put(270,12){1}\put(300,12){0}
      }
      \end{picture}
  \end{center}
    \caption{Top: The generalized current $\eta^{(3)}_2(x)$ is shown in red letters.
    According to (\ref{eq:etaH}) and (\ref{eq:Hmin}), it is evaluated as $\min(2-\alpha_1,\beta_1)$, where 
    $\alpha_1$ and $\beta_1$ are numbers just under and above it, respectively.
    Bottom: Similarly, the generalized current $\eta^{(2)}_3(x)$ is defined using $\min(3-\alpha_1,\beta_1)$.
    One can see on this example that the two currents densities are in fact equal:
    $\eta^{(3)}_2(x)=\eta^{(2)}_3(x)$.
    }
    \label{fig:eta}
\end{figure}
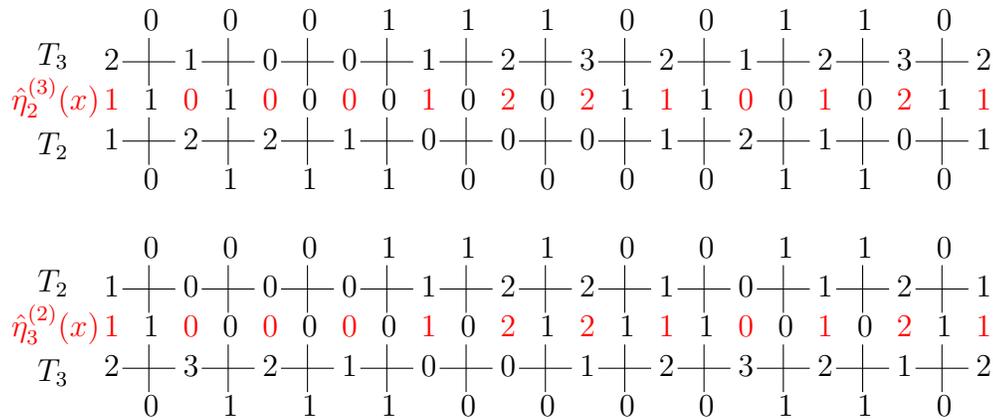

Let us denote by $\hat \eta^{(l)}_j(x)$ the microscopic
operator measuring the current
associated to the $j^{\rm th}$ energy, under the time evolution $T_l$, and at position $x$.
To define this operator microscopically in a given state $s$ of the BBS we need to consider two successive time evolutions $s \rightarrow T_l(s) \rightarrow T_iT_l(s)$,
as illustrated in Fig.~\ref{fig:eta}.
Consider the carrier inducing the first $T_l$ and let $u(x)=(u_0,u_1)$ be its state at position $x$,
where $u_0$ and $u_1$ are the numbers of empty space and balls in it, respectively.
Similarly, let $u'(x) = (u'_0, u'_1)$ be the state of another carrier for the second time evolution $T_i$ at position $x$. 
By the definition of the carrier capacity $u_0+u_1=l$ and $u'_0+u'_1=i$.
Now ${\hat \eta}^{(l)}_i(x)$ of the state $s$ is defined by ${\hat \eta}^{(l)}_i(x) = \min(u'_0,u_1)$ (red integers in Fig.~\ref{fig:eta}).
By extending the argument in Sec.~2.2 in \cite{KMP20}, it can be shown that these generalized currents are in fact symmetric in the two capacities $l$ and $i$:
$\hat{\eta}^{(l)}_i(x) = \hat{\eta}^{(i)}_l(x)$, and
$\hat{\eta}^{(l)}_\infty(x)  = j^{(l)}(x)$ and ${\hat \eta}^{(1)}_k(x)=$ local term for the energy $E_k$.
We refer the reader to \ref{app:eta_def} for more details about the construction of these currents.

The mean values of these energy currents
are expressed in terms of $\eta^{(l)}_j$ (\ref{etadef}) as
\begin{align}\label{eta5}
\eta^{(1)}_j = \left<\hat\eta^{(1)}_j\right>=\sum_{k \ge 1}\min(j,k) \rho_k,
\quad 
\eta^{(l)}_j = \left<\hat\eta^{(l)}_j\right>=\sum_{k \ge 1} \min(j,k)\rho_k v^{(l)}_k,
\end{align}
where $v^{(l=1)}_k=1$ is used.
The mean ball current (\ref{eq:jkj}) corresponds to $\eta^{(l)}_\infty$, and the ball density is $\eta^{(1)}_\infty$.
Comparing (\ref{eta5}) and  (\ref{etadef}), we see that the index $l$ allows interpolating between the 
density and the current, utilizing the variety of time evolutions in the system. 

Now consider the time-averaged correlation
\begin{align}
  {\mathcal C}^{m,l,n}_{i,j}= \lim_{t_n \rightarrow \infty}
  \frac{1}{t_n} \int_0^{t_n} ds\sum_x\left< \hat\eta^{(m)}_i(x,s)\hat\eta^{(l)}_j(0,0)\right>^c,
\end{align}
which is also expected to coincide with the limit
\begin{align}
  {\mathcal C}^{m,l,n}_{i,j}= \lim_{t_n \rightarrow \infty}
  \sum_x\left< \hat\eta^{(m)}_i(x,t_n)\hat\eta^{(l)}_j(0,0)\right>^c
  \label{eq:Cmln}.
\end{align}
In the definitions above we have denoted the time variable by $t_n$ to emphasize that the time evolution from time 0 to time $t_n$ is computed with $T_n$.
We however conjecture that, for sufficiently large $n$ this quantity becomes independent of $n$, and we set
\begin{align}\label{CCdef}
C^{l,m}_{i,j} = \lim_{n \rightarrow \infty}
{\mathcal C}^{m,l,n}_{i,j}.
\end{align}
This property is easy to check
if at least one of the indices $m,l,i,j$ is equal to one. Using the symmetry between the upper and lower indices in $\hat\eta^{(l)}_i$,
the index which is equal to one can be moved to an upper position. 
The correlation (\ref{eq:Cmln}) can then be formulated
using, say, $\sum_x \hat\eta^{(1)}_i(x)$, which is the conserved energy $E_i$.
The time evolution therefore drops and  ${\mathcal C}^{m,l,n}_{i,j}$
appears to be independent of $n$ in such a case.
As we will see in Sec.~\ref{ssec:drude_numerics}, in more general cases the numerical simulations support the fact that ${\mathcal C}^{m,l,n}_{i,j}$
is independent of $n$ for $n\geq \min(m,l)$.

The family of generalized correlations $C^{l,m}_{i,j}$ includes three quantities we have defined previously:
\begin{align}\label{cDR}
c_2= C^{1,l}_{\infty, \infty}, \quad D = C^{l,l}_{\infty,\infty}, 
\quad
R = C^{1,1}_{\infty,\infty}.
\end{align}
As seen here, the superscripts $l,m$ in (\ref{CCdef}) 
are restricted to $1$ (for density) or $l$ (for current) in the original setting.
However, keeping them general reveals an interesting symmetry of the problem as we see below.
Admitting the validity of the previous TBA argument leads to
\begin{align}
C^{l,m}_{i,j} &= \sum_{k=1}^\infty \frac{\partial \eta^{(l)}_i}{\partial \bar{}\epsilon_k} 
\frac{\partial \eta^{(m)}_j}{\partial \bar\epsilon_k} 
\frac{1+e^{\bar\epsilon_k}}{\sigma_k}
\\
&=\sum_{k \ge 1}\rho_k\sigma_k(\rho_k+\sigma_k)v^{(i)}_kv^{(j)}_kv^{(l)}_k v^{(m)}_k.
\label{eq:ClmijTBA}
\end{align}
It tells that $C^{l,m}_{i,j}$ is {\em completely symmetric in the four indices}.

The symmetry relating $C^{l,m}_{i,j}$ to $C^{i,m}_{l,j}$ and $C^{l,j}_{i,m}$
derives from the equality $\hat \eta_i^{(l)}=\hat \eta_l^{(i)}$.
For example,  the ball density correlation $R = C^{1,1}_{\infty,\infty}$ coincides with 
$C^{\infty, \infty}_{1,1}$ which is the soliton current correlation
under the time evolution $T_\infty$.
In this particular case the symmetry can be checked directly at the microscopic level.\footnote{To this end one compares the
$1-$energy density (or soliton density) $e_1(x,t)$ at times $t$ and $t+1$ (according to the $T_\infty$ evolution) to deduce the associated
current $j_1(x,t)$ satisfying the discrete continuity equation
$e_1(x,t+1)-e_1(x,t)+j_1(x+1/2,t)-j_1(x-1/2,t)=0$. Doing so
one realizes that the soliton current is identical to the ball density, up to a half lattice spacing shift:
$j_1(x,t)=n(x-1/2,t)$. This explains the equality between their correlations
and provides an explicit check of the general symmetry mentioned above.} 

The symmetry relating $C^{l,m}_{i,j}$ to $C^{m,l}_{j,i}$
follows from the fact that, in the correlation
$\left< \hat\eta^{(m)}_i(x,s)\hat\eta^{(l)}_j(0,0)\right>^c$
one can replace  $(x,s)$ by $(-x,-s)$. The possibility
to exchange, say, the indices $l$ and $m$ in  $C^{l,m}_{i,j}$
is apparent in the TBA result (\ref{eq:ClmijTBA}) but it does not follow
from a simple microscopic symmetry.
It will be interesting to seek such an enhanced symmetry 
among the transport characteristics also 
in other integrable models admitting the GHD approach.

We note that the family of characteristics
$C^{m,l}_{\infty, \infty}\,(m=1,2,\ldots,l)$
 interpolating $c_2$ and $D$ in (\ref{cDR}) are all expressible as a finite sum 
similar to  (\ref{eq:Dl}): 
\begin{align}
C^{m,l}_{\infty, \infty} = \sum_{k < m}W_k v^{(l)}_k(v^{(m)}_k-v^{(m)}_m)
+ c_2 v^{(m)}_m.
\end{align}

When $i,j,l,m$ are {\em all finite},
$C^{l,m}_{i,j}$ is evaluated as a rational function by means of (\ref{sum9}) with $a=z$ as
\begin{align}
C^{l,m}_{i,j} = \sum_{k<r}\rho_k\sigma_k(\rho_k+\sigma_k)v^{(i)}_kv^{(j)}_kv^{(l)}_kv^{(m)}_k
+ \frac{z^r(1-z)^4(1+z^{2r+1})}{(1+z)^3(1-z^r)^2(1-z^{r+1})^2}
v^{(i)}_iv^{(j)}_jv^{(l)}_lv^{(m)}_m,
\end{align}
where $r=\max(i,j,l,m)$.

\subsection{Numerical calculation of the Drude weights}
\label{ssec:drude_numerics}
\begin{figure}[H]
  \begin{center}
    \includegraphics[width=0.7\linewidth]{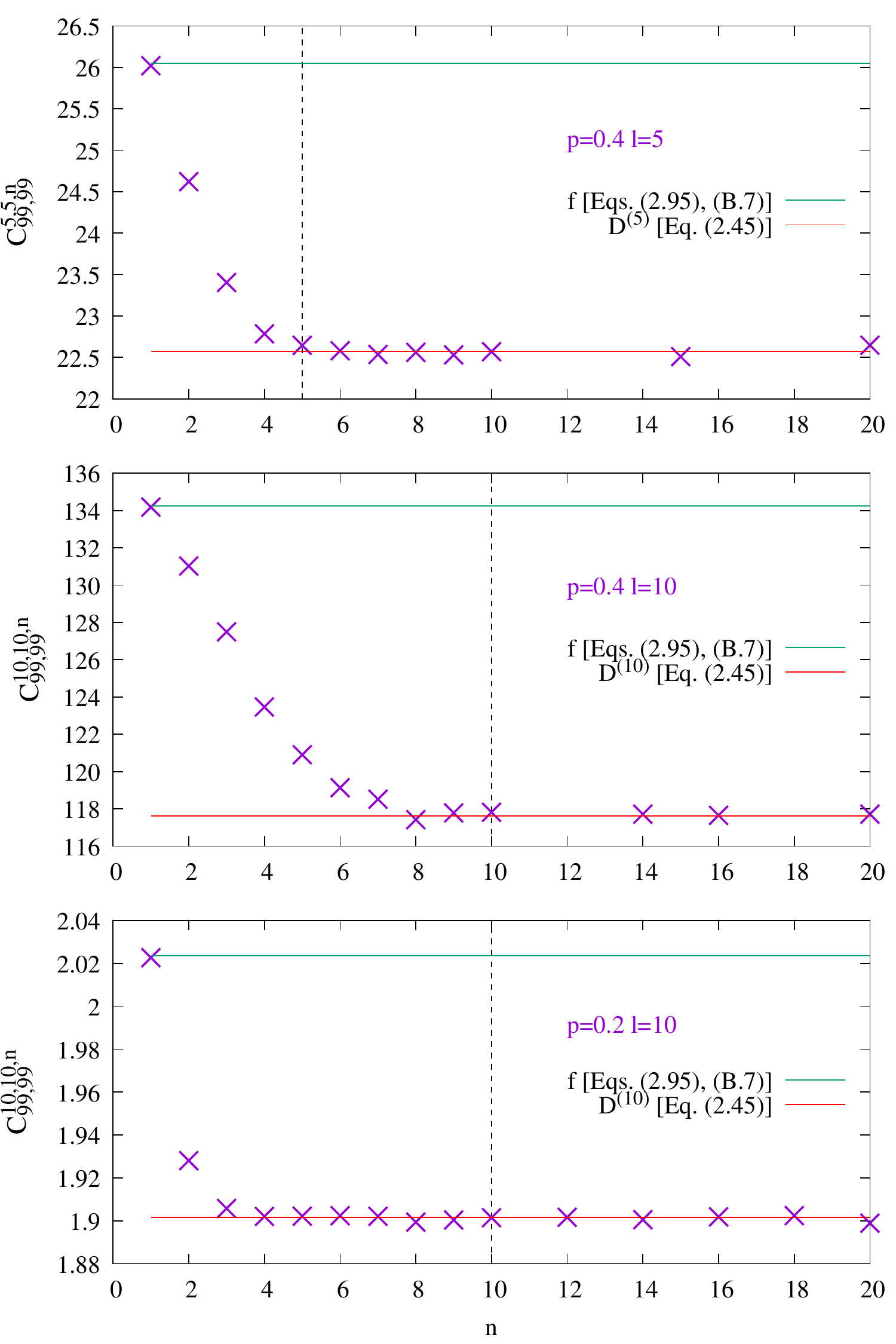}
    \caption{Numerical results for the Drude weight
    $\mathcal{C}^{l,l,n}_{99,99}$ (\ref{eq:Cmln}) for different values of the ball densities ($p=0.4$ and $p=0.2$), two values of $l$
    and different values of the parameter $n$ characterizing the time evolution.
    The simulations were carried out on a periodic system of size $L=60000$,
    with time $t_n=1000$ and $N_{\rm samples}=5\,10^5$ random initial configurations (top and middle panel)
    and with $N_{\rm samples}=8\,10^6$ in the bottom panel.
    The red horizontal lines represent the TBA result (\ref{dsum}) and the green lines
    represent (\ref{eq:conjectured_f}).
    }\label{fig:drude}
  \end{center}
\end{figure}

\begin{figure}[H]
  \begin{center}
    \includegraphics[width=0.7\linewidth]{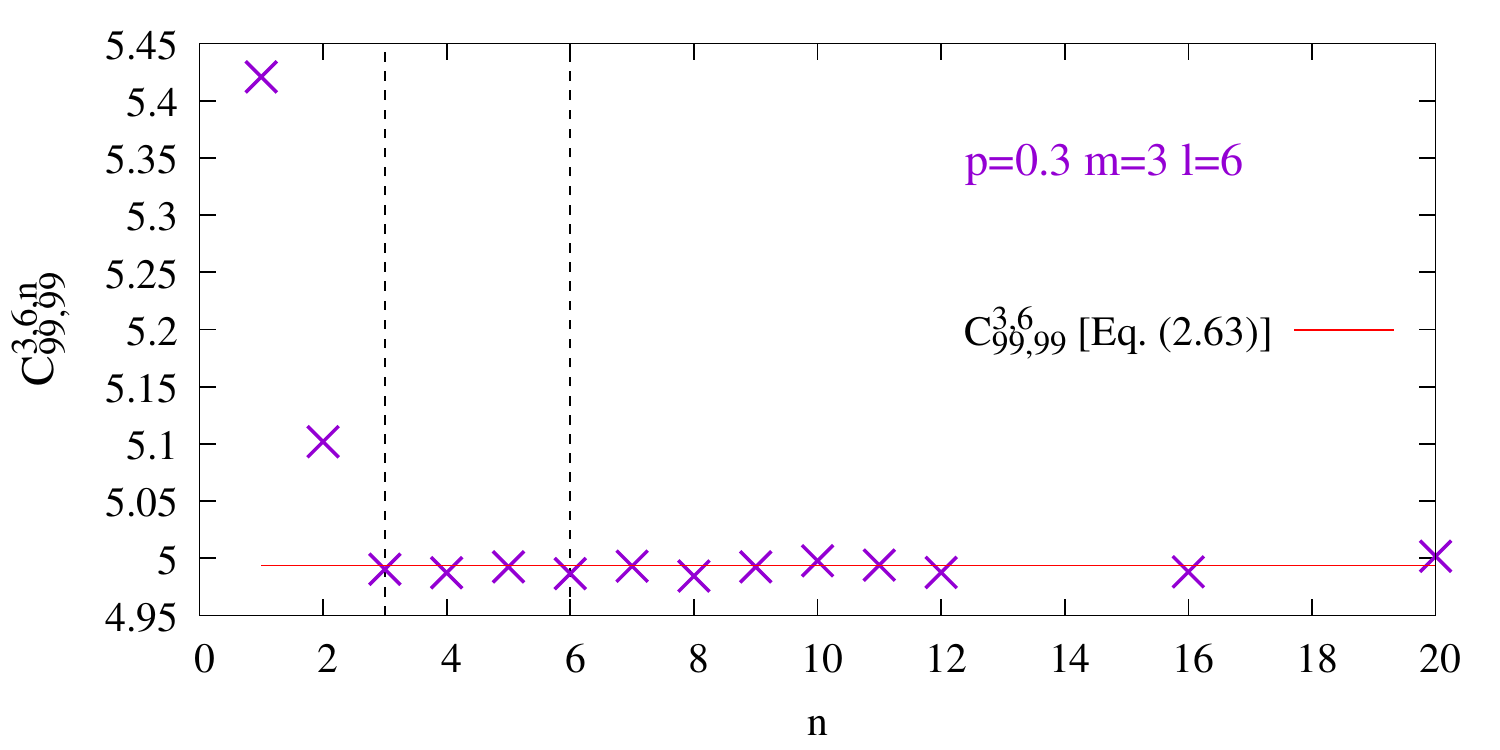}
    \caption{Numerical results for ${\mathcal C}^{3,6,n}_{99,99}$ (\ref{eq:Cmln}) plotted as a function of $n$. The data suggests that ${\mathcal C}^{m,l,n}_{\infty,\infty}$ coincides with the TBA result (\ref{eq:ClmijTBA}) for $C^{m,l}_{\infty, \infty}$ when $n\geq \min(m,l)$.
    Simulations performed with (periodic) system size $L=30000$, time $t_n=1000$ and $1.5\,10^6$ random initial conditions with ball density $p=0.3$.}\label{fig:drude2}
  \end{center}
\end{figure}

\begin{figure}[H]
  \begin{center}
    \includegraphics[width=0.7\linewidth]{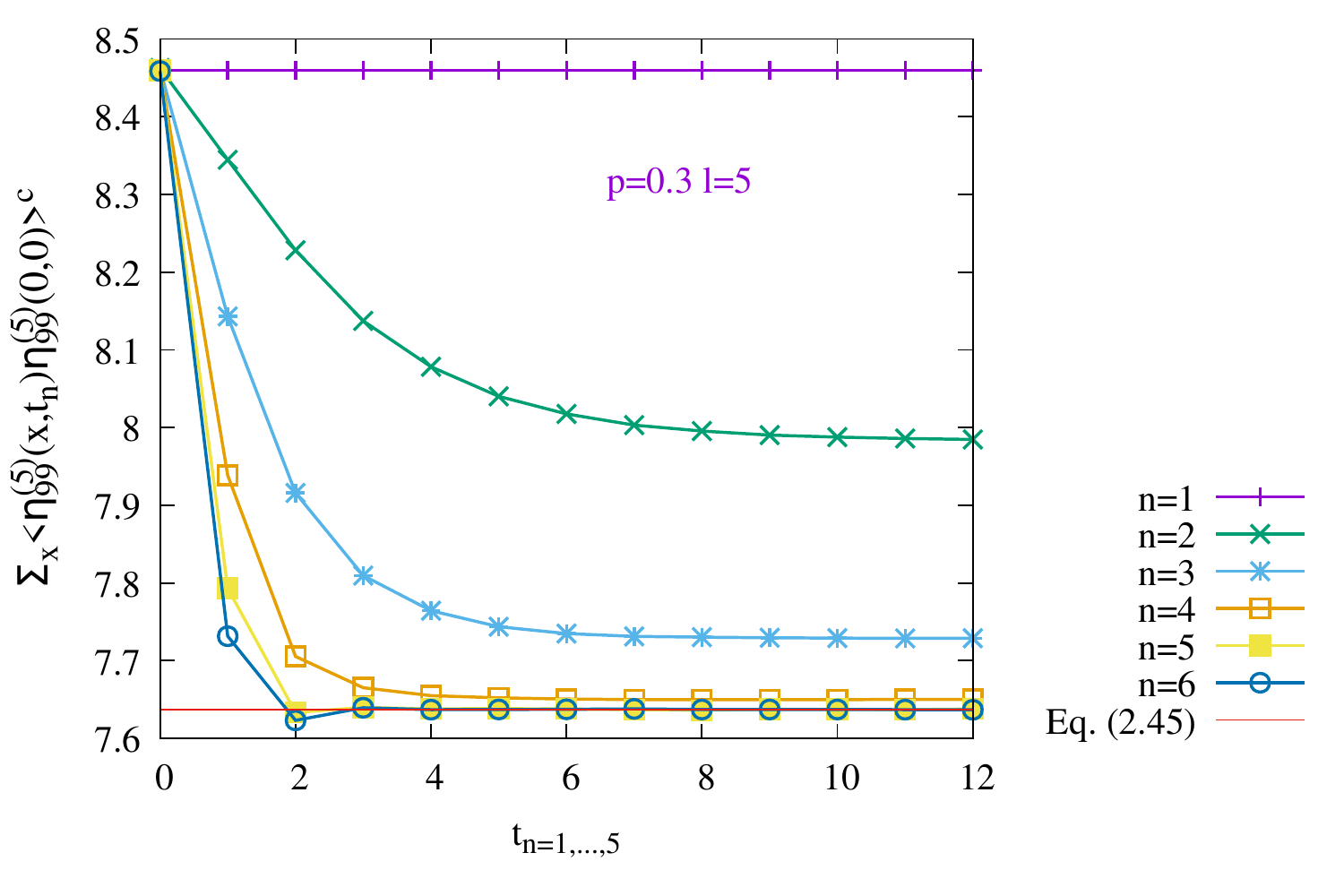} 
    \caption{Correlation
    $\sum_x\langle \eta^{(5)}_{99}(x,t_n)\eta^{(5)}_{99}(0,0)\rangle^c$
    plotted as a function of time for different time evolutions $T_n$.
    Since $T_1$ is a simple translation the dynamics is trivial and the correlation is independent of time (purple curve, $n=1$). For $n\geq2$ the dynamics is however nontrivial, and the correlation relaxes to a finite value at sufficiently long times.
    For $n\geq 5$ this long-time limit coincides with the TBA value (\ref{dsum})
    (horizontal red line).
    Simulations performed with (periodic) system size $L=20000$, and $36\,10^6$ random initial conditions with ball density $p=0.3$.}
    \label{fig:drude_t}
  \end{center} 
\end{figure}

In this section we present some direct numerical calculations
of the correlation ${\mathcal C}^{m,l,n}_{i,j}$. 
The results are displayed in Figs.~\ref{fig:drude}, \ref{fig:drude2} and \ref{fig:drude_t}.

Fig.~\ref{fig:drude} illustrates the dependence of ${\mathcal C}^{m,l,n}_{i,j}$ on the dynamical parameter $n$ in the case $m=l$ and $i=j=99$ (almost $\infty$). Since $T_1$ is a simple translation
we have $\hat\eta^{(l)}_i(x,t_1)=\hat\eta^{(l)}_j(x-t_1,0)$
and (\ref{eq:Cmln}) reduces to an equal-time correlation when $n=1$
(no long-time limit). This equal-time current-current correlation
is computed in Sec.~\ref{ssec:f} and the numerical data for $n=1$
agree with the analytical result. For general values of $n$
we can compare the numerical results for ${\mathcal C}^{l,l,n}_{\infty,\infty}$
with the Drude weight ${\mathcal C}^{l,l}_{\infty,\infty}$, and the 
data plotted in Fig.~\ref{fig:drude} suggest that we have ${\mathcal C}^{l,l,n}_{\infty,\infty}=C^{l,l}_{\infty, \infty}$ for $n\geq l$. We are however unaware of a method to obtain ${\mathcal C}^{l,l,n}_{\infty,\infty}$ in the intermediate
regime $1<n<l$. 

Fig.~\ref{fig:drude2} illustrates the dependence on the dynamical parameter $n$ in a case where $m \ne l$ (and still $i=j=99$). For $n=1$ the correlation
${\mathcal C}^{m,l,1}_{\infty,\infty}$ is again an equal-time correlation between two generalized currents. 
For larger values of the dynamical parameter $n$ the data displayed in Fig.~\ref{fig:drude2} suggest that we have ${\mathcal C}^{m,l,n}_{\infty,\infty}=C^{m,l}_{\infty, \infty}$ for $n\geq \min(m,l)$, even though we have no proof. 

Finally,  Fig.~\ref{fig:drude_t} illustrates the convergence of
current-current correlations
as a function of time and for different time evolutions. This figure shows that it converges to ${\mathcal C}^{l,l,n}_{\infty,\infty}$  after only a relatively short relaxation time. In this particular example ($l=5$ and ball density $p=0.3$) the relaxation time appears to decrease when $n$ increases. As for the long-time limit of the correlator, it coincides with the TBA expression (\ref{eq:ClmijTBA}) only for $n\geq l=5$, in agreement with the conjectured proposed above.

If the long-time limit of a current-current correlation (that is, the Drude weight) is subtracted, the time integral of the remaining part of the correlation is an element of the Onsager matrix, which characterizes the diffusive corrections to the hydrodynamics~\cite{D19,de_nardis_correlation_2022}. We see from Fig.~\ref{fig:drude_t} that such Onsager coefficient would be dominated by a few short-time value of the correlator. The detailed study of the Onsager matrix in the BBS is  left for future studies.\footnote{The diffusive broadening of density steps in a domain-wall problem was studied analytically and numerically for the BBS in \cite{KMP20,kuniba_generalized_2021}.}

In the hydrodynamic projection scheme the Drude weights and its generalizations (\ref{eq:Cmln}) depend on the projections of the current operators onto the space spanned by the conserved quantities. The energies $E_i$ are conserved whatever the dynamical parameter $n$
and their contribution to the long-time limit of the current-current correlations was included in (\ref{eq:ClmijTBA}), via the pseudoenergy variables. However, the numerical observation that 
 ${\mathcal C}^{m,l,n}_{i,j}$
 can be different from (\ref{eq:ClmijTBA}) suggest that $T_n$ may not always be
 sufficiently mixing inside each energy sector fixed by the $\{E_i\}$ data, in particular for small $n$.  This is obvious for $n=1$ (no mixing at all), but as a less trivial example consider two solitons of sizes $p$ and $p'$ under the evolution $T_{n}$, with $n\leq p<p'$.
 Because the effective soliton velocities (\ref{eq:veff}) are independent
 of the soliton size when they are larger than $n$  the mean distance between two big solitons will be almost constant in time (up to small fluctuations  due to the collisions with smaller and slower solitons). This may be a hint why some minimal value of $n$ seems required for a current-current correlation to converge to the TBA result (\ref{eq:ClmijTBA}).

\subsection{Flux Jacobian and related matrices}
\label{ssec:flux_jacobian}

In this section we present concrete forms of the flux Jacobians 
corresponding to the commuting family of time evolutions in our BBS. 
We will also describe the relations with the matrices of 2nd cumulants and the Drude weights,
as well as their generalization $C^{l,m}_{i,j}$.

\subsubsection{Normal modes}
In BBS, the role of the conserved charge $q_i$ and 
the current $j_i$ in the flux Jacobian [eq.~(26) of \cite{D19}] are played by 
$\eta^{(1)}_i$ and $\eta^{(l)}_i$ if the time evolution is taken as $T_l$.
From [eq.~(3.9), \cite{KMP20}], 
$\eta^{(1)}_i$ is related to the hole density as $\sigma_i = 1-2 \eta^{(1)}_i$.
The associated conserved currents for the time evolution $T_l$ are given by  $\sigma_i v^{(l)}_i$. 
Therefore, we have the equations of conservation:
\begin{align}
\partial_{t_l}\sigma_i+\partial_x(\sigma_i v^{(l)}_i)=0.
\label{conserv}
\end{align}
We use the fact that ${\sigma}_i $ and $v^{(l)}_i $ are functions of $y_j$ given by  
$\sigma_{i}=\sum_{k=1}^\infty G_{ik}$ 
and $\sigma_i v^{(l)}_i=(G\kappa^{(l)})_i = (GM)_{il}/2$.
Their partial derivatives with respect to $y_j$ are calculated as
\begin{align}
{\partial \sigma_i\over \partial y_j} &
= -\sum_k y^{-1}_j \frac{\partial G_{ik}}{\partial \bar\epsilon_j} 
= -\sum_k y^{-1}_j 2 \sigma_j v^{(i)}_j y_j G_{jk}
= -2\sigma_j^2 v^{(i)}_j = -(GM)_{ij}\sigma_{j},
\label{elem0}\\
{\partial (\sigma_i v^{(l)}_i)\over \partial y_j}&
= -\frac{1}{2}\sum_k y^{-1}_j\frac{\partial G_{ik}}{\partial \bar\epsilon_j}M_{kl}=  -\sum_k y^{-1}_j \sigma_j v^{(i)}_j y_j G_{jk}M_{kl}\nonumber \\
&
= -2\sigma^2_j v^{(i)}_j v^{(l)}_j = -(GM)_{ij}\sigma_{j}v^{(l)}_j,
\label{elem}
\end{align}
where (\ref{df1}) is used.
Substituting them into (\ref{conserv}), we deduce that $y_j$ are normal modes [eq.~(5.20), \cite{KMP20}]:
\begin{align}
\partial_{t_l}y_j+v^{(l)}_j\partial_x y_j =0.
\label{normal}
\end{align}

\subsubsection{Family of flux Jacobians}
Let us introduce the infinite dimensional matrices
\begin{align}\label{VVS}
V = (v^{(i)}_j)_{i,j \ge 1}, \quad V^{(l)}_d = \text{diag}(v^{(l)}_1,v^{(l)}_2, \ldots), 
\quad {\hat \sigma} = \text{diag}(\sigma_1, \sigma_2, \ldots),
\end{align}
where the effective velocity is related to the hole density  [eq.~(4.19), \cite{KMP20}] as
\begin{align}\label{vf}
v^{(l)}_i = \sum_{k=1}^{\min(l,i)}\frac{\sigma_l}{\sigma_{k-1}\sigma_k}
\end{align}
with $\sigma_0=1$.
We therefore have an explicit expression of the hole currents
\begin{align}\label{eq:j_hole_sigma}
  j_i^{(l)\rm hole}=\sigma_i v^{(l)}_i = \sum_{k=1}^{\min(l,i)}\frac{\sigma_i\sigma_l}{\sigma_{k-1}\sigma_k}
  \end{align}
in terms of the hole densities $\sigma_i$. In the GHD context such an expression is an equation of state [(\cite{D19}), eq. (22)]. 
The results (\ref{elem0}) and (\ref{elem}) are expressed  in matrix forms as 
\begin{align}\label{gmm}
\frac{\partial \sigma}{\partial y} = -GM{\hat \sigma},
\qquad
\frac{\partial (\sigma v^{(l)})}{\partial y} = -GMV^{(l)}_d{\hat \sigma}.
\end{align}
If we use $\sigma_i$ as variables to express the currents, (\ref{conserv}) is written as
\begin{align}
\partial_{t_l}\sigma_i+\sum_{j=1}^\infty A^{(l) j}_i\partial_x\sigma_j=0.
\label{conserv2}
\end{align}
The matrix $A^{(l)}=(A^{(l) j}_{i})_{i,j \ge 1}$ (row $i$, column $j$) is the flux Jacobian, where 
\begin{align}\label{Ass}
A^{(l) j}_i&={\partial {(\sigma_i v_i^{(l)})}\over \partial\sigma_j} 
= \sum_k\frac{\partial (\sigma_iv^{(l)}_i)}{\partial y_k}\frac{\partial y_k}{\partial \sigma_j}.
\end{align}
In terms of matrices in  (\ref{gmm})  this is equivalent to
\begin{align}
A^{(l)} &= GMV^{(l)}_d{\hat \sigma} (GM{\hat \sigma})^{-1}
= GMV^{(l)}_d M^{-1}G^{-1} = MG^t V^{(l)}_d(G^t)^{-1}M^{-1},
\label{vsig}
\end{align}
where $GM = M G^t$ is used.
Note from (\ref{eq:GM_sym}) and (\ref{eq:GM}) that $GM = 2{\hat \sigma}V^t = 2V{\hat \sigma}$.
Thus the above result is also expressed as 
\begin{align}\label{avvv}
A^{(l)} = V V^{(l)}_d V^{-1},\quad \text{i.e.}\quad 
\sum_{k=1}^\infty A_i^{(l) k} v^{(k)}_j &= v^{(l)}_j v^{(i)}_j\qquad 
(\forall i,j, l \in \Z_{\ge 1}).
\end{align}
This confirms the fact that the effective velocities $v^{(l)}_j\,(1 \le j \le l)$ 
are the eigenvalues of $A^{(l)}$ [eq.~(30) of \cite{D19}].

We remark that our flux Jacobians form a commuting family:
\begin{align}
[A^{(l)}, A^{(l')}]=0\qquad (A^{(1)}= 1).
\end{align}
This comes from the fact that the matrix $V$ which diagonalizes $A^{(l)}$ does not depend on $l$ (only the eigenvalues depend on $l$). This property
reflects the commutativity $[T_l, T_{l'}]=0$ of the time evolutions in BBS. 

Substitution of (\ref{vf}) into the first expression of (\ref{Ass}) yields 
\begin{align}
A_i^{(l) j} 
= \sum_{k=1}^{\min(l,i)}\frac{\partial}{\partial \sigma_j}
\Bigl(\frac{\sigma_i \sigma_l}{\sigma_{k-1}\sigma_k}\Bigr).
\label{fjac}
\end{align}
A little inspection of this shows that 
$A^{(l)}$ 
has the block structure whose top left block is of size $l \times l$,
the top right one is zero
and the bottom right one is $v^{(l)}_l$ times the identity matrix of infinite size:
\begin{align}
A^{(l)} &= \begin{pmatrix} \mathscr{A}^{(l)} & 0 \\ \mathscr{U}  & v^{(l)}_l \mathrm{I} \end{pmatrix},
\\
\mathscr{A}^{(l)}   &= \Bigl(\frac{\partial(\sigma_i v^{(l)}_i)}{\partial \sigma_j}\Bigr)_{1 \le i,j \le l},
\quad 
\mathscr{U} = \begin{pmatrix}
\sigma_{l+1} u_1 & \sigma_{l+1} u_2 \, \ldots  & \sigma_{l+1} u_l 
\\
\sigma_{l+2} u_1 & \sigma_{l+2} u_2  \,\ldots  & \sigma_{l+2} u_l 
\\
 & \cdots &
 \end{pmatrix},
 \quad
  u_j = \frac{\partial v^{(l)}_l}{\partial \sigma_j},
\end{align}
For instance $\mathscr{A}^{(3)}$ reads as
\begin{align}
\mathscr{A}^{(3)} = \left(
\begin{array}{ccc}
 0 & 0 & 1 \\
 -\frac{(\sigma_2+1) \sigma_3}{\sigma_1^2} & \frac{\sigma_3}{\sigma_1} & \frac{\sigma_2+1}{\sigma_1} \\
 -\frac{(\sigma_2+1) \sigma_3^2}{\sigma_1^2 \sigma_2} & -\frac{\sigma_3 (\sigma_1+\sigma_3)}{\sigma_1 \sigma_2^2} & \frac{\sigma_1+2 \sigma_2 \sigma_3+2 \sigma_3}{\sigma_1
   \sigma_2} \\
\end{array}
\right).
\end{align}

\subsubsection{Covariance and Drude matrices}
\label{sssec:cov_drude_mat}

Let us introduce the matrices 
$C=(C_{i,j})_{i,j \ge 1}$,
$B^{(l)}=(B^{(l)}_{i,j})_{i,j \ge 1}$ 
and $D^{(l)} = (D^{(l)}_{i,j})_{i,j \ge 1}$
whose elements are special cases of $C^{l,m}_{i,j}$ in (\ref{eq:ClmijTBA}):
\begin{align}
C_{i,j} &=  C^{1,1}_{i,j} = \sum_p\rho_p\sigma_p(\rho_p+\sigma_p)v^{(i)}_pv^{(j)}_p,
\label{Cij}
\\
B^{(l)}_{i,j} &= C^{1,l}_{i,j} = \sum_p\rho_p\sigma_p(\rho_p+\sigma_p)v^{(l)}_pv^{(i)}_pv^{(j)}_p,
\label{Bij}\\
D^{(l)}_{i,j} & = C^{l,l}_{i,j} = \sum_p \rho_p\sigma_p(\rho_p+\sigma_p)(v^{(l)}_p)^2v^{(i)}_pv^{(j)}_p.
\label{Dij}
\end{align}
The quantities $C^{(l)}_{i,j}$ and $D^{(l)}_{i,j}$ 
are static covariance and  Drude weights [eqs.(158), (163),  of \cite{D19}].

By using (\ref{avvv}) it is easy to see
\begin{align}\label{ABCD}
A^{(l)} C= CA^{(l) t}, \quad B^{(l)} = A^{(l)}C, \quad D^{(l)}= \bigl(A^{(l)}\bigr)^2C
= A^{(l)}CA^{(l) t}.
\end{align}
In fact, the first relation is a special case of 
$\sum_kA_i^{(l) k} {\tilde C}_{k,j} = \sum_k {\tilde C}_{i,k}A_j^{(l) k}$ 
for ${\tilde C}_{i,j} = \sum_p w_p v^{(i)}_pv^{(j)}_p$ with arbitrary parameters $w_k$.
The equality of the coefficient of each $w_p$ follows directly from (\ref{avvv}).

Recall the matrix form of the flux Jacobian in (\ref{vsig}), i.e., 
\begin{align}
A^{(l)} = M (1+\hat{y}M)^{-1} V^{(l)}_d(1+\hat{y}M) M^{-1}.
\label{AMM}
\end{align}
Introduce further diagonal matrices
\begin{align}
\hat{\rho} = \mathrm{diag}(\rho_1,\rho_2, \ldots),
\qquad
{\hat g} =  1+ {\hat y} = \mathrm{diag}(1+y_1, 1+y_2,\ldots).
\end{align}
Then the formula 
\begin{align}
C_{i,j} &= \sum_p \rho_p(1+y_p)(\sigma_p v^{(i)}_p)(\sigma_pv^{(j)}_p)
=\frac{1}{4}\sum_p (GM)_{pi}\rho_p(1+y_p)(GM)_{pj}
\end{align}
derived from (\ref{eq:GM_sym}) and (\ref{eq:GM})  is interpreted as 
$C= \frac{1}{4}MG^t {\hat \rho}{\hat g}GM$ due to $GM = MG^t$.  
Applying (\ref{ABCD}) to this and (\ref{AMM}) we find 
\begin{align}
C &=\frac{1}{4}M(1+\hat{y}M)^{-1} \hat{\rho}{\hat g} (1+M\hat{y})^{-1} M,
\label{Cus}
\\
B^{(l)} &= \frac{1}{4}M (1+\hat{y}M)^{-1}V^{(l)}_d \hat{\rho}{\hat g} (1+M\hat{y})^{-1} M,
\label{Bus}
\\
D^{(l)} &= \frac{1}{4}M (1+\hat{y}M)^{-1}(V^{(l)}_d)^2 \hat{\rho}{\hat g} (1+M\hat{y})^{-1}M.
\label{Dus}
\end{align}
The factor $\frac{1}{4}$ originates in the coefficient 2 in $M_{ij}=2\min(i,j)$, which is a non-essential
artifact.
The results (\ref{AMM}) and (\ref{Cus}) -- (\ref{Dus}) agree with 
[eqs.(165)--(168), \cite{D19}].
\footnote{
The correspondence with eqs.~(127), (149), (157) etc. in \cite{D19} as follows:
$\rho_p(k) \leftrightarrow \rho_k$,
$\rho_s(k) \leftrightarrow \sigma_k$,
$n(k) \leftrightarrow (1+1/y_k)^{-1}$,
$f(p) \leftrightarrow (1+ y_p)^{-1}$,
$E'(p) \leftrightarrow \min(l,p)$ (for $T_l$ BBS dynamics),
$h_i(p) \leftrightarrow \min(i,p)$,
$(E')^{dr}(p) \leftrightarrow \sigma_pv^{(l)}_p$,
$h^{dr}_i(p) \leftrightarrow \sigma_p v^{(i)}_p$,
$1^{dr}(p) \leftrightarrow \sigma_p$,
$\int\frac{dp}{2\pi} \leftrightarrow \sum_{p=1}^\infty$.
}
We remark a natural generalization
\begin{align}
C^{(l,m)} := (C^{l,m}_{i,j})_{i,j \ge 1}  = A^{(l)}C\bigl(A^{(m)}\bigr)^t  (= C^{(m,l)})
\end{align}
covers $C=C^{(1,1)}, B^{(l)} = C^{(l,1)}=C^{(1,l)}$ and 
$D^{(l)}= C^{(l,l)}$.

\subsection{Equal-time current-current correlations}
\label{ssec:f}
As a slight digression, we consider here the spatially integrated current
\begin{equation}
  J = \sum_x j(x,0)
\end{equation}
and its variance
\begin{equation}
  f=L^{-1} \left(\left< J^2\right>-\left< J\right>^2\right)
  =\sum_x \left< j(x,0) j(0,0)\right>^c,
  \label{eq:fdef}
\end{equation}
where the factor $L^{-1}$ insures that $f$ is finite in the thermodynamic limit.
The r.h.s of (\ref{eq:fdef}) is analogous to the current-current correlation (\ref{eq:Djj}) which defines the Drude weight, except for the important fact it is an equal-time correlation.

It is possible to compute $f$ using a transfer matrix approach, as explained in \ref{sec:TM}.
$f$ of course differs from the Drude weight, but  the numerical values turn out to be close (see Tab.~\ref{tab:drude}). Both diverge
when approaching the half-filled limit $p\to 1/2$ at $l=\infty$, and both are equal to $p(1-p)$ for $l=1$ (in which case the BBS dynamics reduces to translations). We are however unaware of a simple way to obtain $f$ using TBA.
The fact that $f\geq D$ is a general property,  which follows from the Cauchy-Scwharz inequality associated to a suitable scalar product between observables (see (3.14) of \cite{spohn_interacting_2018}).

\begin{table}[H]
  \begin{center}\begin{tabular}{|c|c|c|c|}
      \hline
      Ball density $p$ & $l$ & $D$ (\ref{eq:Dtba}) (\ref{dsum}) & $f$ (\ref{eq:fdef}) (\ref{eq:conjectured_f})\\
      \hline
      $p$              & 1   & $p(1-p)$            &$p(1-p)$            \\
      0.2              & 2   & 0.6383506           &0.672498            \\
      0.2              & 5   & 1.744263            &1.857532            \\
      0.2              & 10  & 1.901627            &2.023662            \\
      0.4              & 2   & 1.606536            &1.790640            \\
      0.4              & 5   & 22.570957           &26.04813            \\
      0.4              & 10  & 117.6194            &134.2497390         \\
      \hline
    \end{tabular}
  \end{center}\caption{
    Drude weight $D$ for a few values of the ball density $p$ and carrier capacity $l$.
    For comparison the total current variance $f$ is also given.
    For $l=1$ the BBS dynamics reduces to translations.
    In such a case the spatially integrated current is trivially independent of time
    and $f=D$ (compare (\ref{eq:Djj}) and (\ref{eq:fdef})). 
  }\label{tab:drude}
\end{table}

\section{Large deviation function}
\label{sec:ldf}
\setcounter{footnote}{0}

We are interested in the fluctuations of a quantity which, on average, grows linearly with time. In such a case the large deviation function is a useful tool to characterize the fluctuations. In general the large deviation function contains the information  about all the cumulants of the fluctuating quantity (full counting statistics).\footnote{See however \cite{krajnik_anisotropic_2021,krajnik_exact_2022} for an example where is this not the case.}
It also describes the rate at which large fluctuations occur \cite{touchette_large_2009}. The large deviation function is in general a difficult quantity to obtain, but some important progress has been made recently in the context of integrable systems \cite{myers_transport_2020,doyon_fluctuations_2020}, where a general method to obtain the large deviation function associated to the transport of a conserved charge has been constructed. In this section we apply these ideas to the case of the BBS, where the calculations simplify considerably.

\subsection{General method}
We want to compute the large deviation associated to the number $N_t$ of balls transferred from the left to the right during time $t$, in some simple GGE stationary state.\footnote{We refer the reader to \ref{sec:2TGGE} for the large deviation function
associated to the joint distribution of the number of transferred balls and the number of transferred solitons.}
By definition, if the principle of large deviation is obeyed, we have in the large time limit:
\begin{equation}
  F(\lambda)=\lim_{t\to \infty}\frac{1}{t}\ln\left[ \langle e^{\lambda{N_t}}\rangle_\beta \right].
\end{equation} Or, in a more compact way:
\begin{equation}
  e^{tF(\lambda)}\sim\langle e^{\lambda{N_t}}\rangle_\beta,
  \label{delN}
\end{equation}
where ${N_t}=\sum_{s=0}^t {j(0,s)}  $ is the number of balls passing through the origin during the time interval $[0,t]$. In the above expression the expectation value $\left< f(t)\right>_\beta$ of an operator $f$ at time $t$ means:
\begin{equation}
  \langle f(t) \rangle_\beta={{\sum_c f(c(t)) e^{-\beta Q(c)} }
  \over \sum_c  e^{-\beta Q(c)}}
\end{equation}
where $c(t)$ is the ball configuration $c$ evolved up to time $t$, and $Q(c)$ is (conserved) total number of balls in $c$.
The parameter $\beta$ describing the i.i.d. state is related to the ball fugacity by $e^{-\beta}=z$.
We focus here on a single conserved quantity (the total number of balls), but the approach can be generalized to a GGE with several $\beta^i$ coupled to several conserved energies, and a function $F$ of several variables $\lambda^i$. See \ref{sec:2TGGE} for a two-$\beta$ case.
Expanding $F$ in powers of $\lambda$ gives access to the scaled cumulants $c_n$:
\begin{equation}
  F(\lambda)=\sum_{n=1}^\infty \frac{\lambda^n}{n!}c_n \;\;{\rm and}\;\;
  c_n=\lim_{t\to \infty} \frac{1}{t}\left<(N_t)^n\right>^c_\beta.
\end{equation}
From the expression above it appears clearly that $F$ is well-defined only if all the cumulants $ \left<(N_t)^n\right>^c_\beta$ have the same $t$-linear scaling.
It should be noted that there are some integrable models where the above relation is not obeyed \cite{krajnik_breakdown_2021,krajnik_exact_2022}.

Taking the derivative with respect to $\lambda$, we have:
\begin{equation}
  {dF(\lambda)\over d \lambda}=
  \lim_{t\to \infty}{1\over t}{{\langle N_t e^{\lambda {N_t}}\rangle_\beta}
    \over {\langle e^{\lambda N_t}\rangle_\beta}}
  \label{eq:F=J}
\end{equation}
where we assume that the above large-time limit exists.

Let us denote by $Q^L_t$ and $Q^R_t$ the charge in the left and right halves of the system at time $t$.
We have $Q^L_t+Q^R_t=Q$, independent of time. Thus,
\begin{equation}
  N_t =  Q^L_0- Q^L_t =  Q- Q_0^R- Q_t^L.
\end{equation}
The relation (\ref{eq:F=J}) can be written
\begin{equation}
  {dF(\lambda)\over d \lambda}=
  \lim_{t\to \infty}{1\over t}{{\langle {{N_t}} e^{-\lambda ( Q_0^R+ Q_t^L)}\rangle_{ \beta -\lambda}}
    \over \langle e^{-\lambda ( Q_0^R+ Q_t^L)}\rangle_{ \beta - \lambda}},
  \label{eq:F=J_2}
\end{equation}
where the factor $\exp(\lambda Q)$ has been converted into a shift $ \beta \to  \beta -  \lambda$.

\begin{figure}[h]
  \begin{center}
    \includegraphics[width=0.35\linewidth]{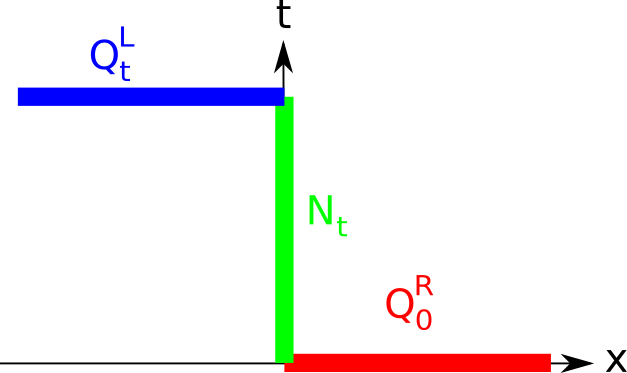}
    \caption{The density operators $n(x>0,t=0)$ appearing in $Q^R_0$ are uncorrelated from the current operators $j(x=0,t>0)$ in $N_t$. In the same way,
      the density operators $n(x<0,t>0)$ appearing in $Q^L_t$ are uncorrelated from $N_t$ and from $Q^R_0$.
    }\label{fig:schema}
  \end{center}
\end{figure}

In the i.i.d. states we consider the equal-time connected correlation of two local operators, $\left<O(x_1,t)O(x_2,t)\right>^c$ vanish if $x_1\ne x_2$. But since the ball propagation only takes place in the $x>0$ direction,
we also know, by causality, that a two-time correlation of the form $\left<O(x_1,t_1)O(x_2,t_2)\right>^c$ vanishes if $(x_1-x_2)(t_1-t_2)<0$.
This means that the local terms appearing in $Q_0^R+Q_t^L$ (acting at $x>0$ and at time $0$ or at $x<0$ and time $t>0$), and the terms in  $ N_t$ ($x=0$ and $t>0$) are uncorrelated.
The term $e^{-\lambda ( Q_0^R+ Q_t^L)}$ thus decouples from $ N_t$ (Fig.~\ref{fig:schema})
and cancels between the numerator and denominator of (\ref{eq:F=J_2}). This leads to
\begin{equation}
  {dF(\lambda)\over d \lambda}=
  \lim_{t\to \infty}{1\over t} \langle {N_t} \rangle_{ \beta-\lambda}.
  \label{eq:F=J_3}
\end{equation}
The state defined by $\langle \cdots \rangle_{ \beta - \lambda}$ is an i.i.d. stationary state, so the expectation value of the currents is independent of time, and we get
\begin{equation}
  {dF( \lambda)\over d \lambda}=
  {\langle  j(0) }\rangle_{ \beta-\lambda}
  \label{F=j}
\end{equation}
where $ j(0)$ is the ball current at the origin (and time zero).

The equation (\ref{F=j}) is closely related to the so-called ``flow equation'' \cite{doyon_fluctuations_2020, myers_transport_2020}. As explained in these two works,
for a general integrable system the derivative ${dF(\lambda)\over d \lambda}$ of the SCGF is the expectation value $\left< j\right>_{\tilde \beta(\lambda)}$ of the current
in a modified GGE state parameterized with modified inverse temperatures $\tilde \beta^i(\lambda)$. These inverse temperatures are determined by integrating the flow equation $\frac{d\tilde\beta^i}{d\lambda}=-{\rm sign}(A(\lambda))_{i^*}^i$
from the initial condition where $\tilde\beta^i(0)=\beta^i$ are the parameters of the original GGE (single parameter $e^{-\beta}=z$ in the i.i.d. case we consider).
The index $i$ labels the conserved charges and $i^*$ is the index of the particular charge for which the SCGF is computed.
We compute here the SCGE associated to the number of balls, so  $i^\ast=\infty$. The matrix $A_{ik}(\lambda)={\partial \left<j_i\right>_{\beta(\lambda)}}/{\partial q_k}$, is the flux Jacobian (see Sec.~\ref{ssec:flux_jacobian}), it appears when linearizing the Euler equations. $A(\lambda)$ is defined by the derivatives of the current densities with respect to the charges densities, evaluated in the $\lambda$-modified state. Its eigenvalues are the effective velocities $v_i$ of the normal hydrodynamical modes (the soliton velocities in the BBS case). ${\rm sign}(A)$ is defined as the matrix with the same eigenvectors as $A$ but replacing each eigenvalue $v_i$ by ${\rm sign} (v_i)=\pm 1$. Here comes a drastic simplification in the BBS case: all the soliton velocities are positive and, therefore, ${\rm sign}(A)={\rm Id}$ and the flow equation becomes $\frac{d\tilde\beta^i}{d\lambda}=-\delta_{i^*,i}$.
Since in the present case $i^\ast=\infty$ the index $i$ is thus also fixed to $i=\infty$.
It follows that the $\lambda-$modification of the state is a simple shift of the parameter $\beta$ associated to the ball density. The  $\lambda-$modified state remains i.i.d.  but with $\tilde\beta(\lambda)=\beta-\lambda$.

\subsection{Scaled cumulants generating function}
In the r.h.s of (\ref{F=j}) we need to consider the mean ball current in a modified i.i.d. state with parameter $\beta'=\beta-\lambda$, and hence
a modified ball fugacity $z'=e^{-\beta+\lambda}$.
\begin{equation}
  {{dF}(\lambda)\over d\lambda}=j(e^{-\beta+\lambda}).
\end{equation}
Writing the integration we obtain
\begin{equation}
  F(\lambda)=\int_0^\lambda j(e^{-\beta+\lambda'})d\lambda' =\int_z^{z e^{\lambda}} \frac{dz'}{z'}j(z'). \label{eq:Fjb}
\end{equation}
Reintroducing explicitly the parameter $l$ of the dynamics (carrier capacity) and using the explicit form of the ball current (\ref{eq:jb}),
the integration in (\ref{eq:Fjb}) can be carried out explicitly and gives
\begin{equation}
  {F^{(l)}(\lambda)}=\ln\left({1-(ze^\lambda)^{l+1}\over 1-ze^\lambda}\right)-\ln\left({1-z^{l+1}\over 1-z}\right).
  \label{eq:Fl}
\end{equation}
As a sanity check we can compute its second derivative at $\lambda=0$:
\begin{equation}
  c_2=\left. \frac{d^2F^{(l)}}{d\lambda^2}\right|_{\lambda=0}
\end{equation}
and we recover (\ref{eq:c2b}). 

The \ref{sec:2TGGE} presents
some generalization of the above results to a two-temperature GGE, with one parameter coupled to the total number of balls and another one to the total number of solitons. In particular a generalization of (\ref{eq:Fl}) is given in (\ref{F2}).

It is useful to consider the Legendre transform of $F$ with respect to $\lambda$, the so-called large-deviation rate function:
\begin{equation}
  G(j)=j \lambda(j) - F(\lambda(j)),
  \label{eq:G}
\end{equation}
where $\lambda(j)$ is a solution of
\begin{equation}
  F'(\lambda(j))=j.
\end{equation}
The probability to observe a transferred charge $N=j\,t$ at time $t$
is then
\begin{equation}
  P_t(N)\sim \exp\left( - t \,G(N/t)\right).
\end{equation}

\begin{figure}
  \begin{center}
    \includegraphics[width=0.8\linewidth]{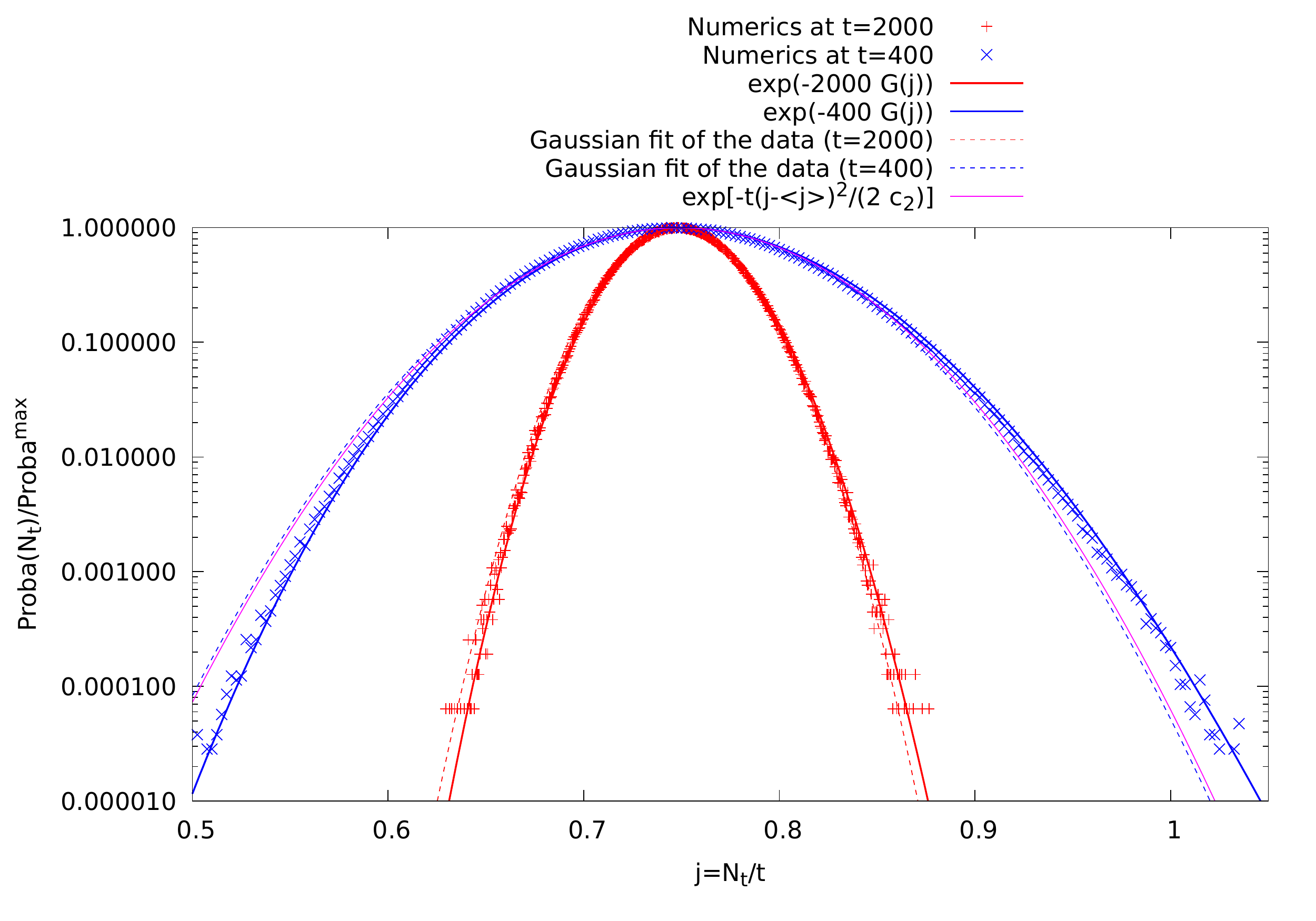}
    \caption{
      Probability distribution of the number $N_t$ of balls having crossed the origin between $t=0$ and $t=400$ (blue), and between $t=0$ and $t=2000$ (red).
      Ball density $p=0.3$ and capacity $l=10$. The numerical data at $t=400$ are obtained using $6\,10^6$ random initial conditions and the data at $t=2000$ involve $2\,10^6$  random initial conditions. These results are compared with the Gaussian
      obtained by doing a least-square fit  of the data (dashed lines). Such a fit has two free parameters: mean and variance.
      The data at $t=400$ are
      also compared to the Gaussian $\exp[-t(j-<j>)^2/(2 c_2)]$ (magenta line), without any adjustable parameter, where $<j>=0.7490144$ is the exact mean value of the current (\ref{eq:jb}) and $c_2=1.3016578$ is the exact second cumulant (\ref{eq:c2b}). Finally, the data are compared to the exact large deviation rate function (continuous lines) $\exp(-tG(N/t))$.  At $t=400$ it is possible to
      see that the numerical data departs from a simple Gaussian and are instead consistent with the theoretical prediction involving the large deviation rate function $G$ (\ref{eq:G}).
    }
    \label{fig:num}
  \end{center}
\end{figure}

$G(j)$ is defined for  $0<j<l$, since the maximum possible value of the ball current is $l$. It  is a convex function obeying $G(0)=\ln(({1-z^{l+1})/( 1-z)})$, $G(l)=G(0)-l \ln(z)$ reaching its maximum zero at the mean current $ j=F'(0)$.
The function $G$ is represented in Fig.~\ref{fig:num}, in the vicinity of its maximum, for one particular case ($p=z/(1+z)=0.3$ and $l=10$).

\subsection{$l=\infty$}

In the limit $l\to\infty$ the SCGF (\ref{eq:Fl}) simplifies to
\begin{equation}
  {F^{(\infty)}(\lambda)}=\ln\left({1-z \over 1-ze^\lambda}\right).
  \label{F_i}
\end{equation}
The interval where $F$ is defined is $\lambda \in [-\infty,\beta]$, which is equivalent to $ze^\lambda<1$ (recall that $z=e^{-\beta}$). For $\lambda \to -\infty$ the current $j=F'(\lambda)$ tends to zero, and for $\lambda \to \beta$ we have instead $j=F'(\lambda)\to +\infty$. The full range of physical values for the ball current is covered by $F'$, as it should.
A generalization of (\ref{F_i}) to a two-temperature GGE is given in (\ref{eq:Finf_2TGGE}).

\begin{figure}
  \begin{center}
    \includegraphics[width=0.8\linewidth]{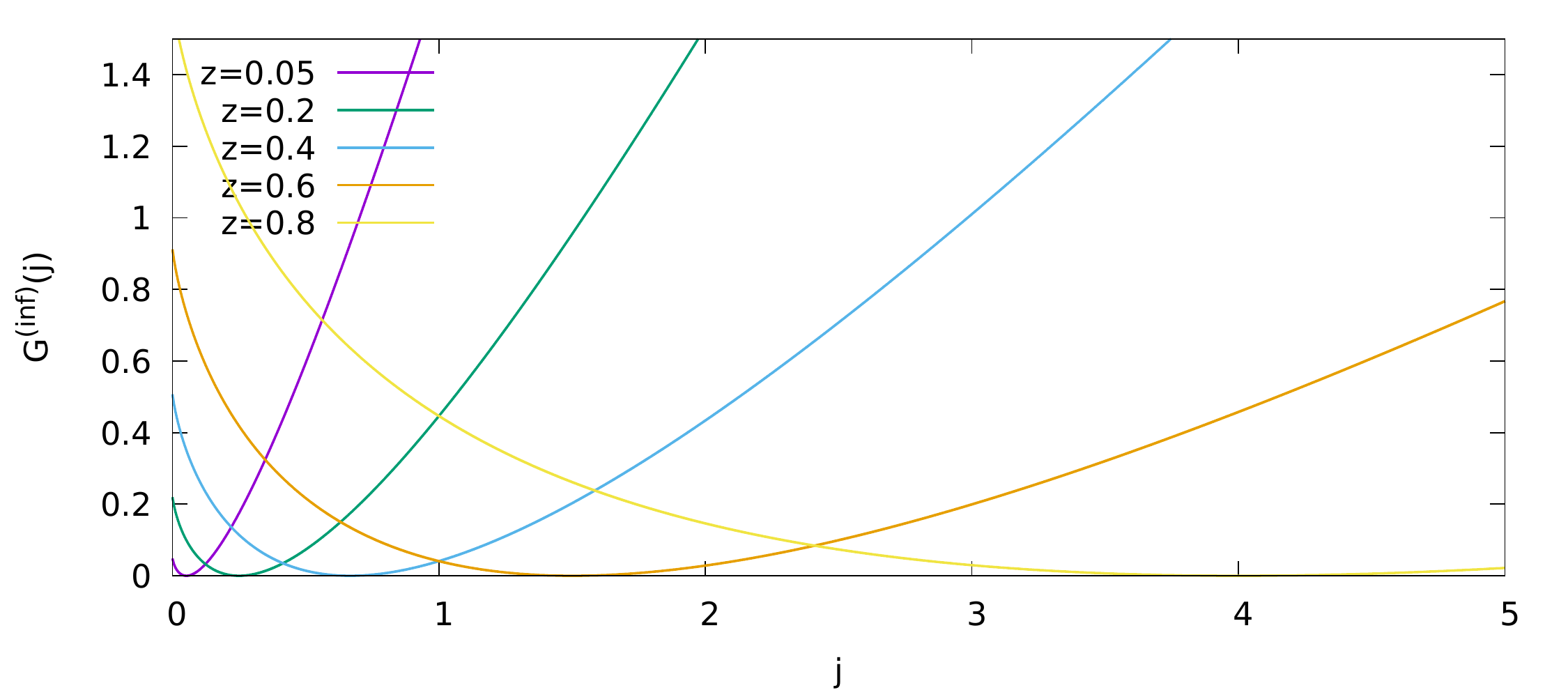}
    \caption{Large-deviation rate function $G^{(l=\infty)}(j)$ plotted as a function of the current $j$, for different values of the ball fugacity $z$. The mean current (minimum of $G$) and the second cumulant $c_2$ diverge when $z\to 1^-$.
    }
    \label{fig:G}
  \end{center}
\end{figure}

In that case the large-deviation rate function can be obtained explicitly:
\begin{equation}
  G^{(\infty)}(j)=-\ln(1-z)-j \ln(z)-(1+j)\ln(1+j)+j\ln(j).
\end{equation}
This function is plotted in Fig.~\ref{fig:G} for a few different values of the ball fugacity. The generalization to a two-temperature GGE is given in (\ref{G2re}). When $z\to 1^-$ $F^{(\infty)}(\lambda)$ becomes singular in $\lambda=0$ and the associated cumulants diverge. For instance: $c_2=z/(1-z)^2$ at $l=\infty$. This can be interpreted as a phase transition when the ball density approaches $1/2$. Some other properties of this transition were discussed in \cite{LLP17}.

To conclude this section
we note that the large deviation principle for the energies $E_k$ 
in i.i.d. states was studied for a multicolor BBS by regarding the history of  carriers going through the states as a Markov process \cite{KL20}. In that previous study the system size is the variable which plays a role similar to the role played by time here.

\section{Conclusions}

We have computed analytically several quantities related to the current and density fluctuations in stationary i.i.d. states. We have obtained the SCGF associated to the number $N_t$ of balls crossing the origin during time $t$, from which all the cumulants can be extracted. The Legendre transform of this function could be compared with the probability distribution of $N_t$ extracted from numerical simulations.
This is one of the very few interacting and deterministic models where the SCGF could be computed exactly
(\cite{krajnik_exact_2022} is another recent example). 

The Drude weights -- defined as the long-time limit of a spatially integrated current-current correlation -- could also be obtained using TBA combined with  hydrodynamical projection. Explicit analytical expressions for the Drude weights could be compared successfully with numerical simulations. 

The existence of a family of commuting time evolutions is an important property of integrable systems,
although implementing them in actual systems can be cumbersome in practice. 
The BBS is a distinguished example of integrable cellular automata where the
all commuting time evolutions have a simple and neat implementation.
Exploiting these time evolutions a set of new generalized currents correlations (or generalized Drude weights) was constructed and shown to enjoy unexpected symmetry relations. Some of these symmetry relations can be explained at the microscopic level, while others only emerge in the long-time limit.
We observed that the numerical results for the current correlations at long times
coincide with the TBA results only for certain time evolution, namely $T_n$ with sufficiently large $n$. This suggests that $T_n$ for small $n$ is insufficiently mixing
and the observed correlations in this regime still escape our understanding.
Understanding these results and reconciling them with the hydrodynamic projection ideas would certainly deserve further study.

We also stress that a number of these results could be generalized to
a larger family of (non i.i.d.) GGE states with two temperatures, respectively coupled to the number of balls and to the number of solitons.

It is quite remarkable that so many explicit formulae could be obtained
for nontrivial quantities related to long-distance and long-time limit of correlations in such an out-of-equilibrium interacting problem. These could prove to be useful to compare BBS with other models, either integrable or non-integrable, and to shed some light about fundamental questions like the emergence of hydrodynamics.

\section{Acknowledgements}

V.~P. thanks E.~Ilievski, Z.~Krajnik, T.~Prosen and J.~Schmidt for numerous discussions and collaboration on related subjects. We also acknowledge
the DRF of CEA for providing us with CPU time on the supercomputer TOPAZE at CCRT.

\appendix
\setcounter{footnote}{0}

\section{Correlation sum rules}\label{sec:sr}
We derive a sum rule connecting density-density correlations to current-current ones. The argument is directly inspired from \cite{mendl_current_2015}.
We consider the following quantity
\begin{equation}
  \sum_{x=-\infty}^\infty f(x)\left( n(x,t)-n(x,0)\right),
  \label{eq:f}
\end{equation}
where $n$ is the ball density operator, and $f$ some test function defined on the lattice sites.
Next we introduce the charge of the region $]-\infty,x]$:
\begin{align}
  N^\langle(x,t)&=\sum_{y=-\infty}^x n(y,t) \\
  n(x,t)&=N^\langle(x,t)-N^\langle(x-1,t)
\end{align}
(\ref{eq:f}) can be rewritten as
\begin{align}
  &\sum_{x=-\infty}^\infty f(x)\left( n(x,t)-n(x,0)\right) \nonumber\\
  &=\sum_{x=-\infty}^\infty f(x)\left( N^\langle(x,t)-N^\langle(x-1,t) - N^\langle(x,0)+N^\langle(x-1,0)\right) \\
  &=\sum_{x=-\infty}^\infty \left(f(x)-f(x+1)\right)\left(N^\langle(x,t)-N^\langle(x,0)\right)\\
  &=-\sum_{x=-\infty}^\infty \left(f(x)-f(x+1)\right)\int_0^t j(x+1,s) ds \\
  &=\sum_{x=-\infty}^\infty \left(f(x)-f(x-1)\right)\int_0^t j(x,s) ds.
  \label{eq:f1}
\end{align}
The current can also be used to write
\begin{equation}
  n(0,t)-n(0,0)=\int_0^t \left(j(0,s)-j(1,s)\right) ds.
  \label{eq:n0}
\end{equation}
We then multiply (\ref{eq:f1}) by (\ref{eq:n0})
\begin{align}
  &\sum_{x=-\infty}^\infty f(x)\left( n(x,t)-n(x,0)\right)\left(n(0,t)-n(0,0)\right) \nonumber\\
  &=\sum_{x=-\infty}^\infty \left(f(x)-f(x-1)\right)\int_0^t j(x,s) ds
  \int_0^t \left(j(0,s')-j(1,s')\right) ds'
\end{align}
and take the connected average:
\begin{align}
  &\sum_{x=-\infty}^\infty f(x)\left<\left( n(x,t)-n(x,0)\right)\left(n(0,t)-n(0,0)\right)\right>^c \nonumber\\
  &=\sum_{x=-\infty}^\infty \int_0^t \int_0^t ds ds' \left(f(x)-f(x-1)\right)\left<  j(x,s)\left(j(0,s')-j(1,s')\right)\right>^c \\
  &=\sum_{x=-\infty}^\infty \int_0^t \int_0^t ds ds' \left(f(x)-f(x-1)\right)\left<  \left(j(x,s)-j(x-1,s)\right)j(0,s')\right>^c
\end{align}
where, in the last equality, we have used the translation invariance of the current-current correlator.
This can finally be rewritten
\begin{align}
  &\sum_{x=-\infty}^\infty f(x)\left<\left( n(x,t)-n(x,0)\right)\left(n(0,t)-n(0,0)\right)\right>^c \nonumber\\
  &=\sum_{x=-\infty}^\infty \int_0^t \int_0^t ds ds' \left(2f(x)-f(x+1)-f(x-1)\right) \left<  j(x,s) j(0,s')\right>^c.
  \label{eq:sum_rule}
\end{align}

\section{Transfer matrix calculation for the current fluctuations}
\label{sec:TM}
This section presents a transfer matrix calculation of the fluctuations of the current (\ref{eq:fdef}). It is an extension of the calculation
presented in the Appendix E of \cite{KMP20}.

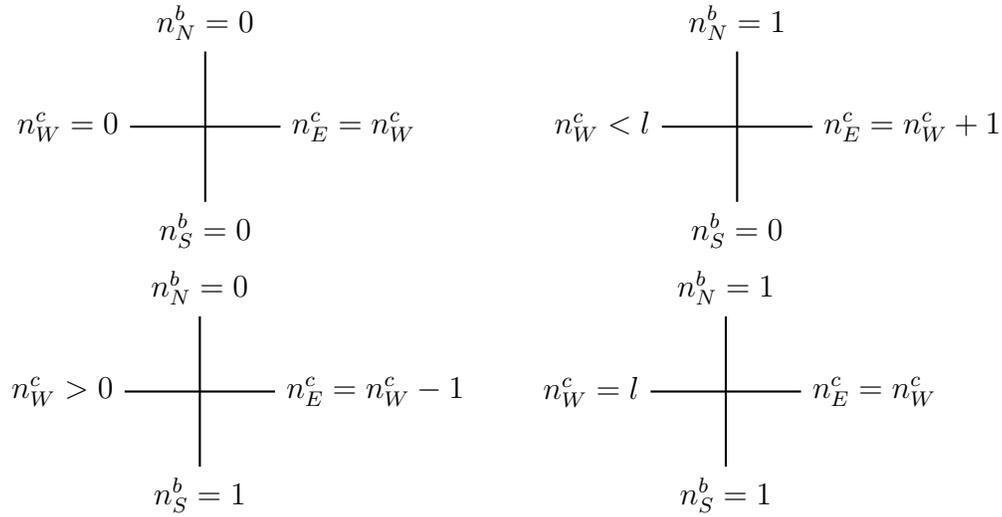
\begin{figure}[!h]
  \centering
  \hspace*{0.8cm}
  \begin{tikzpicture}[scale=1]
    \draw[thick] (-1,0) node[left]{$n^c_W=0$}--(1,0) node[right]{$n^c_E=n^c_W$};
    \draw[thick] (0,-1)node[below]{$n^b_S=0$} --(0,1)node[above]{$n^b_N=0$};
  \end{tikzpicture}
  \hspace*{1.3cm}
  \begin{tikzpicture}[scale=1]
    \draw[thick] (-1,0) node[left]{$n^c_W<l$}--(1,0) node[right]{$n^c_E=n^c_W+1$};
    \draw[thick] (0,-1)node[below]{$n^b_S=0$} --(0,1)node[above]{$n^b_N=1$};
  \end{tikzpicture}\\
  \begin{tikzpicture}[scale=1]
    \draw[thick] (-1,0) node[left]{$n^c_W>0$}--(1,0) node[right]{$n^c_E=n^c_W-1$};
    \draw[thick] (0,-1)node[below]{$n^b_S=1$} --(0,1)node[above]{$n^b_N=0$};
  \end{tikzpicture}
  \hspace*{0.5cm}
  \begin{tikzpicture}[scale=1]
    \draw[thick] (-1,0) node[left]{$n^c_W=l$}--(1,0) node[right]{$n^c_E=n^c_W$};
    \draw[thick] (0,-1)node[below]{$n^b_S=1$} --(0,1)node[above]{$n^b_N=1$};
  \end{tikzpicture}
  \caption{Four examples of vertices.}
  \label{fig:vertices}\end{figure}

Here we consider the periodic BBS model on a chain of length $L$. It admits a discrete set  of commuting  evolutions characterized by  a row to row transfer propagator $T_l$ labelled  by a positive integer $l$ representing the capacity of the carrier. The propagator $T_l$ takes the form of a vertex transfer matrix (see Fig.~\ref{fig:vertices}) where the horizontal links can contain up to $l$ balls which are auxiliary variables: $0\le n^c\le l$. The vertical links contain zero or one ball: $0\le n^b\le 1$ which are the BBS variables.
Time flows down, so that at each time step, the south vertical links  are occupied according to the north vertical link configurations. We can see the evolution as the result of the passage of a carrier transporting up to $l$ balls from west to east and updating each vertical link successively passing through the vertices.
If the north vertical link is empty ($n^b_N=0$) and the carrier has at least one ball ($n^c_W>0$), it leaves one of its balls to the south vertical link during the passage ($n^b_S=1$, $n^c_E=n^c_W-1$).
If it does not carry balls ($n^c_W=0$), it passes without changing either the vertical occupation or its load ($n_S=0$, $n^c_E=0$).
If the north vertical link is filled ($n^b_N=1$) and the carrier
carries strictly less than $l$ balls ($n^c_W<l$), it picks up a ball and leaves the south vertical link empty ($n^c_E=n^c_W+1$, $n^b_S=0$).
If it carries $l$ balls ($n^c_W=l$),  it passes without changing either the vertical occupation or its load ($n^b_S=1$, $n^c_E=l$). On a periodic chain, it can be shown (Proposition 5.1 of \cite{IKT12}) that the periodicity condition on the horizontal links uniquely determines the load of the carrier.\footnote{To be precise: it is so when the density is not exactly equal to 1/2.}

As a result, the transfer propagator can be expressed as:
\begin{equation}
  (T_l)^{n^b(t)}_{n^b(t+1)}=\left< n^b(t+1)\right|{\rm Tr}\left(\prod_{k=1}^L \mathbb{L}_k\right)\left| n^b(t)\right>
  \label{propagator}
\end{equation}
where $n_b(t)$ stands for the configuration $\{n^b(x,t),1\le x\le L\}$ and
$\langle n^b_S|\mathbb{L}| n^b_N\rangle= \langle n^b_S|\mathbb{L}| n^b_N\rangle_{n^c_W,n^c_E}$ is a $(l+1)\times (l+1)$ permutation matrix with nonzero matrix elements equal to one whenever the corresponding vertex is allowed.

Due to charge conservation, $n_N+n_W=n_S+n_E$, the ball density $n^b(x,t)$ coincides with the number of balls on the vertical links and the ball current  $j(x,t)=n^c(x,t)$ to the number of balls on the horizontal links.

We can give a fugacity to the balls by inserting the operator $z^{Q}=z^{\sum_{x=1}^L n^b(x,t)}$ which commutes with the propagator. Denote $|0\rangle$ the state obtained by summing all the ball configurations with weight one. The i.i.d. stationary state with ball density $p=z/(z+1)$ is $z^{Q}|0\rangle$.

We can also give a fugacity  to the total current at time $t$ by weighting the horizontal links by $y^{J(t)}=y^{\sum_{x=1}^L j(x,t)}$.
We denote by $T_l(y)$ (or $\mathbb{L}(y)$ for a single site) the modified propagator where we  weight each vertex by $y^{n^c_W}$ instead of $1$, which has the effect to weight each configuration by $y^{J(t)}$.
\begin{equation}  
\begin{picture}(50,80)(-100,-30) 
  \put(-240,0){$\mathbb{L}^a(y)_{i,j} =
  \begin{cases} y^i &  \\
   \phantom{Q} & \\
    0 &
  \end{cases}$}  
  \put(-110,18){
  \put(-20,-2){if}
  \put(-7,-2){$i$}\put(12,13){$a$}\put(33,-3){$j$}\put(12,-16){$a^\ast$}
  \put(0,0){\line(1,0){30}}
  \put(15,-10){\line(0,1){20}}
  \put(44,-2){is in Figure \ref{fig:vertices} ($a^\ast = a+i-j$)}
  \put(-20,-40){otherwise.}
  }
\end{picture}
\end{equation}

Consider the matrix element $Z=\langle 0|z^{Q}y^{J(t)}|0\rangle={\rm Tr}(\mathbb{L}^0(y)+z\mathbb{L}^1(y))^L$ where $\mathbb{L}^0(y),\ \mathbb{L}^1(y)$ are the deformed vertex matrices with $n^b_N=0,1$ respectively ($n^b_S$ is  redundant).

Denoting $\mathbb{L}^z(y)=\mathbb{L}^0(y)+z\mathbb{L}^1(y)$, we have:
\begin{align}
\langle 0|\mathbb{L}^z(y)&=\langle 0|+z\langle 1|
\nonumber\\
\langle n|\mathbb{L}^z(y)&=y^n\langle n-1|+zy^n\langle n+1|\ {\rm if}\ n\ne 0,l
\nonumber\\
\langle l|\mathbb{L}^z(y)&=y^l\langle l-1|+zy^l \langle l|.
\label{matrixelement}
\end{align}
As an example, for $l=3$ we have
\begin{align}
  \mathbb{L}^z(y)&=
  \begin{pmatrix}
   1 & z & 0 & 0 \\
   y & 0 & zy & 0 \\
   0 & y^2 & 0 & zy^2 \\
   0 & 0 & y^3 &zy^3
  \end{pmatrix}.
  \end{align}
The matrix element $\mathbb{L}^z(1)_{\alpha\beta}/(z+1)$ is the probability to have $\beta$ balls on the east link knowing that there are $\alpha$ on the west link. Therefore, the components of the left eigenvector of $\mathbb{L}^z(1)/(z+1)$ with eigenvalue one, $\sum_{k=0}^l z^k\langle k|$, are the unnormalized probabilities for the carrier to contain $k$ balls. The current formula (\ref{eq:jb}) is obtained as the average number of balls in the carrier.

The second cumulant $c_2$ and $f$,  the integrated  current-current connected correlation, are then given by second derivatives of $\ln(Z)$:
\begin{equation}
  c_2=\frac{1}{L}\left(z\frac{\partial}{\partial z}\right)\left(y\frac{\partial}{\partial y}\right)\ln(Z)|_{y=1},
\end{equation}
\begin{equation}
  f=\langle J(0)j(0,0)\rangle^c=\frac{1}{L}\left(y\frac{\partial}{\partial y}\right)^2\ln(Z)|_{y=1}.
\end{equation}

By analyzing the values of $f$ for various values of $l$, it has also been possible to conjecture the following analytical expression for $f$:
\begin{equation}
  \begin{split}
    f=&\frac{z(1+6z+z^2)(1-z^{3l+3})}{(1-z)^4(1-z^{l+1})^3}
    -\frac{24z^{l+5}(1-z^{l-3})}{(1-z)^4(1-z^{l+1})^3}
    -\frac{3z^{l+2}(g_{l+2}+7z g_l + 8z^2 g_{l-2})}{(1-z)^3(1-z^{l+1})^3}
    \\
    & - \frac{(l+1)^2z^{l+1}(g_{l+2}+2l(1+z)g_{l+1}+3zg_l)}{(1-z)(1-z^{l+1})^3},
  \end{split}
  \label{eq:conjectured_f}
\end{equation}
where $g_j = 1+z^j$.

\section{Drude weight and SCGF in a two-temperature GGE}
\label{sec:2TGGE}
The BBS has the conserved quantities 
$E_k=L \sum_{j\ge 1}\min(j,k) \rho_j$ for $k=1,2,\ldots$ \cite{KMP20}.
$E_\infty$ is the total number of balls with which the main text is concerned.
In this appendix we present a partial generalization of the results 
to the two-temperature GGE corresponding to the statistical weight
$e^{-\beta_1 E_1 - \beta_\infty E_\infty}$.
The conserved quantity $E_1$ is the number of solitons.
The inverse temperature $\beta$ in the main text is denoted by $\beta_\infty$ here.

Following \cite{KMP20} (3.20) we parameterize the temperatures by $a,z$ as
\begin{align}\label{qlim}
\mathrm{e}^{\frac{1}{2}\beta_1} 
= \frac{a^{\frac{1}{2}}-a^{-\frac{1}{2}}}
{z^{\frac{1}{2}}-z^{-\frac{1}{2}}},
\qquad
\mathrm{e}^{-\beta_\infty} = z.
\end{align}
The single temperature GGE($\beta_\infty$) corresponds to the limit $a \rightarrow z$.
Derivatives by the temperatures are expressed in terms of $a$ and $z$ as in [eq.~(3.22) and (3.23) in\cite{KMP20}]:
\begin{align}
\frac{\partial}{\partial \beta_1}
&= 
\frac{\partial a}{\partial \beta_1}\frac{\partial }{\partial a}
+\frac{\partial z}{\partial \beta_1}\frac{\partial }{\partial z}
= -\frac{a(1-a)}{1+a}\frac{\partial}{\partial a},
\label{ee1}
\\
\frac{\partial}{\partial \beta_\infty}
&= 
\frac{\partial a}{\partial \beta_\infty}\frac{\partial }{\partial a}
+\frac{\partial z}{\partial \beta_\infty}\frac{\partial }{\partial z}
= 
-\frac{a(1-a)(1+z)}{(1+a)(1-z)}\frac{\partial }{\partial a}
-z\frac{\partial }{\partial z}.
\label{ee2}
\end{align}

Densities, effective velocity and currents have been obtained in \cite{KMP20} as follows
((\ref{sj}) was not included therein):
\begin{align}
\sigma_k &=  \frac{(1-a)(1+a z^k)}{(1+a)(1-az^k)},
\qquad
\rho_k= \frac{az^{k-1}(1-a)(1-z)^2(1+az^k)}
{(1+a)(1-az^{k-1})(1-az^k)(1-az^{k+1})},
\label{roaz}
\\
v^{(l)}_k &= \frac{1+az^{l}}{1-az^{l}}v_{\min(k,l)},
\qquad 
v_k = \frac{1+a}{1-a}k - \frac{2a(1+z)(1-z^k)}{(1-a)(1-z)(1+az^{k})}.
\label{vaz}
\\
&\text{ball current}: \; j^{(l)}_\infty 
= \sum_{k\ge 1} k \rho_k v^{(l)}_k 
= \frac{a(1+z)(1-z^l)}{(1+a)(1-z)(1-az^l)}-\frac{l a z^l}{1-az^l},
\label{bj}
\\
&\text{soliton  current}: \;j^{(l)}_1
= \sum_{k\ge 1} \rho_k v^{(l)}_k
= \frac{a(1-z^l)}{(1+a)(1-az^l)}.
\label{sj}
\end{align}
The results (\ref{bj}) and (\ref{sj}) are the $j=\infty$ and $j=1$ cases of (\ref{eta5}):
\begin{align} 
\eta^{(l)}_j = \frac{a(1+z)(1-z^{\min(j,l)})(1+az^{\max(j,l)})}
{(1+a)(1-z)(1-az^j)(1-az^l)} - \frac{\min(j,l)a(z^j+z^l)}{(1-az^j)(1-az^l)}.
\end{align}
Set $W_i = \rho_i\sigma_i(\rho_i+\sigma_i)v^2_i$ as in the main text.
Then the sum formula (\ref{Asum}) admits the following generalization:
\begin{align}
\sum_{i<l}W_i &= \frac{a(1-a)(1+z)A_l}{(1+a)^3(1-z)(1-az^{l-1})^2(1-az^l)^2},
\\
A_l &= 1+2a(1+a^2)z^{2l-1}+a^4z^{4l-2}+a^2z^{2l-2}(1+8z+z^2)
-a z^{l-1}(2+a^3z^{2l-1})g_l(z)
\nonumber \\
&
- z^{l}(1+2a^3z^{2l-1})g_l(z^{-1})
-a^2z^{l-1}h_l(z)-a^2z^{3l-1}h_l(z^{-1}),
\label{A2}\\
g_l(z) &=  \frac{z+\bigl(1-l(1-z)\bigr)^2}{1+z},
\quad
h_l(z) = (1+l)^2+l^2z+\frac{ (1-2l+z)(3+z+2lz)}{1+z}.
\end{align}
This is easily shown by induction on $l$ and $A_0=A_1=0$.
The above $A_l$ precisely reduces to (\ref{Asum}) in the single temperature case $a=z$.

Define the second cumulant $c_2^{(l)}$ and the Drude weight 
$D^{(l)}$ by (\ref{eq:c2tba_2}) and (\ref{dsum}) with 
$\rho_i, \sigma_i, v^{(l)}_i$ specified in (\ref{roaz})--(\ref{vaz}).
From $v^{(1)}_i=1$ we have 
\begin{align}
D^{(1)} &= \sum_{i\ge 1}W_i = 
\frac{a(1-a)(1+z)A_\infty}{(1+a)^3(1-z)}
=\frac{a(1-a)(1+z)}{(1+a)^3(1-z)},\label{eq:D1_2TGGE}
\\
c^{(l)}_2 
&= \frac{a(1-a)(1+z)^2(1-z^l)(1+a^2z^l)}{(1+a)^3(1-z)^2(1-az^l)^2}
+ \frac{2az(1-z^l)}{(1+a)(1-z)^2(1-az^l)}
\nonumber\\
& \quad - \frac{alz^l}{(1-az^l)^2}
\left(l+\frac{2(1-a)(1+z)}{(1+a)(1-z)}\right) \label{eq:c2l}
\\
&= -\frac{\partial j^{(l)}_\infty}{\partial \beta_\infty}.
\end{align}
For $D^{(l)}$ with general $l$, formally the same formula as (\ref{eq:Dl}) with 
$\frac{z}{(1+z)^2}$ replaced by (\ref{eq:D1_2TGGE}) is valid.
These results reduce to the single temperature case at $a=z$.
Another useful sum formula is
\begin{align}\label{sum9}
\sum_{i\ge r}\rho_i\sigma_i(\rho_i+\sigma_i) &=
\frac{a(1-a)^3(1-z) z^{r-1} \left(1+a^2 z^{2 r-1}\right)}
{(1+a)^3 \left(1-a z^{r-1}\right)^2 \left(1-az^r\right)^2}.
\end{align}

The ball density $p$ in the two-temperature GGE is known to be  $p=\frac{a}{1+a}$ in
[\cite{KMP20} (3.23)].
Unlike (\ref{eq:p1mp}), the result (\ref{eq:D1_2TGGE}) does not coincide with the $p(1-p)$ reflecting the fact that the two-temperature GGE under consideration is {\em not} i.i.d.

It is natural to introduce the joint cumulant generating function
\begin{align}
F(\lambda, \mu) = \lim_{t \rightarrow \infty}\frac{1}{t}
\ln \langle e^{\lambda N_{\infty,t} + \mu N_{1,t}}\rangle_{\beta_\infty, \beta_1},
\end{align}
where the superscript  $l$ is suppressed and 
$N_{\infty,t} = \int_0^t j^{(l)}_\infty(0,\tau)d\tau$ and 
$N_{1,t} = \int_0^t j^{(l)}_1(0,\tau)d\tau$.
By the same argument as before, we have
\begin{align}\label{Fj}
\frac{\partial F(\lambda,\mu)}{\partial \lambda} = j^{(l)}_\infty(\beta_\infty - \lambda, \beta_1-\mu),
\qquad 
\frac{\partial F(\lambda,\mu)}{\partial \mu} = j^{(l)}_1(\beta_\infty - \lambda, \beta_1-\mu).
\end{align}
Let $(\zeta,\alpha)$ be the parameters $(z,a)$ corresponding to 
$(\beta_\infty - \lambda, \beta_1-\mu)$ in the sense of (\ref{qlim}).
Then from (\ref{bj}) and (\ref{sj}) the currents are expressed as
\begin{align}
j^{(l)}_\infty(\beta_\infty - \lambda, \beta_1-\mu) 
&= 
\frac{\alpha(1+\zeta)(1-\zeta^l)}{(1+\alpha)(1-\zeta)(1-\alpha\zeta^l)}
-\frac{l \alpha \zeta^l}{1-\alpha\zeta^l},
\label{bj2}\\
j^{(l)}_1(\beta_\infty - \lambda, \beta_1-\mu)
&= \frac{\alpha(1-\zeta^l)}{(1+\alpha)(1-\alpha\zeta^l)}.
\label{sj2}
\end{align}
Here $(\zeta,\alpha)$ are regarded as functions of $(\lambda, \mu)$ including $(z,a)$ as parameters. 
Namely,  
\begin{align}\label{qlim2}
\frac{\alpha^{\frac{1}{2}}-\alpha^{-\frac{1}{2}}}
{\zeta^{\frac{1}{2}}-\zeta^{-\frac{1}{2}}}
= \mathrm{e}^{\frac{1}{2}(\beta_1-\mu)}  = 
\frac{a^{\frac{1}{2}}-a^{-\frac{1}{2}}}
{z^{\frac{1}{2}}-z^{-\frac{1}{2}}}e^{-\frac{1}{2}\mu}
\qquad
\zeta = \mathrm{e}^{-\beta_\infty+\lambda} = ze^{\lambda}.
\end{align}
They imply the derivative relations similar to (\ref{ee1}) and (\ref{ee2}):
\begin{align}\label{ee3}
\frac{\partial}{\partial \lambda} 
= \frac{\alpha(1-\alpha)(1+\zeta)}{(1+\alpha)(1-\zeta)}\frac{\partial}{\partial \alpha}
+ \zeta \frac{\partial}{\partial \zeta},
\qquad
\frac{\partial}{\partial \mu}
= \frac{\alpha(1-\alpha)}{1+\alpha}\frac{\partial}{\partial \alpha}.
\end{align}
By using (\ref{bj2}), (\ref{sj2}) and  (\ref{ee3}), one can directly check 
the consistency of (\ref{Fj}):
\begin{align}
\frac{\partial  j^{(l)}_\infty(\beta_\infty - \lambda, \beta_1-\mu)}{\partial \mu}
= 
\frac{\partial  j^{(l)}_1(\beta_\infty - \lambda, \beta_1-\mu)}{\partial \lambda}.
\end{align}

The solution to (\ref{Fj}) satisfying $F(0,0) = 0$ is given by 
\begin{align}\label{F2}
F(\lambda, \mu) = \ln\left(\frac{1-\alpha \zeta^l}{1-\alpha}\right)
- \ln\left(\frac{1-a z^l}{1-a}\right),
\end{align}
where $\alpha = \alpha(\lambda, \mu)$ and $\zeta=\zeta(\lambda,\mu) (= ze^\lambda)$ are 
specified as the solution to (\ref{qlim2}) satisfying $\alpha>0$.
Note that $\alpha(\lambda, \mu=0)\vert_{a=z}= ze^\lambda$.
Thus, the result (\ref{F2}) provides a generalization of (\ref{eq:Fl}) 
reproducing the latter as
$F(\lambda, \mu=0)\vert_{a=z} = F^{(l)}(\lambda)$. 

One can check $\left.\frac{\partial^2 F}{\partial \lambda^2}\right|_{\lambda=\mu=0}=c_2^{(l)}$.
The other second order scaled cumulants are given by 
\begin{align}
\lim_{t \rightarrow \infty} \frac{1}{t} \langle N_{\infty,t}N_{1,t}\rangle^c_{\beta_\infty, \beta_1}
&= 
\left.\frac{\partial^2 F}{\partial \lambda \partial \mu}\right|_{\lambda=\mu=0}
=
\frac{a(1-a)\bigl((1+z)(1-z^l)(1+a^2z^l)-(1+a)^2(1-z)lz^l\bigr)}
{(1+a)^3(1-z)(1-az^l)^2},
\\
\lim_{t \rightarrow \infty} \frac{1}{t} \langle N_{1,t}^2\rangle^c_{\beta_\infty, \beta_1}
&= 
\left.\frac{\partial^2 F}{\partial \mu^2}\right|_{\lambda=\mu=0}
=\frac{a(1-a)(1-z^l)(1+a^2z^l)}{(1+a)^3(1-az^l)^2}.
\end{align}

We leave an interesting problem of studying the Hessian of $F$ in relation to 
the convexity of $F$ for a future work. 

When $l=\infty$, (\ref{F2})  in the regime $ze^\lambda<1$ simplifies to 
\begin{align}
F^{(\infty)}(\lambda, \mu) = \ln\left(\frac{1-a}{1-\alpha}\right).
\label{eq:Finf_2TGGE}
\end{align}
This is still a non-trivial function of $\lambda$ and $\mu$ via (\ref{qlim2}).
From (\ref{Fj}) -- (\ref{sj2}) with  $l = \infty$, 
the equation 
$\Bigl(
\frac{\partial F^{(\infty)}(\lambda,\mu)}{\partial \lambda}, 
\frac{\partial F^{(\infty)}(\lambda,\mu)}{\partial \mu}\Bigr)
= (\mathscr{J}_\infty, \mathscr{J}_1)$ 
has the solution
\begin{align}
\lambda_\ast = \ln\left(
\frac{\mathscr{J}_\infty-\mathscr{J}_1}{z(\mathscr{J}_\infty+\mathscr{J}_1)}\right),
\quad
\mu_\ast = \ln\left(\frac{4(1-\mathscr{J}_1)\mathscr{J}_1^3(a+a^{-1}-2)}
{(\mathscr{J}_\infty^2-\mathscr{J}_1^2)(1-2\mathscr{J}_1)^2(z+z^{-1}-2)}\right),
\end{align}
which is deduced from
$\alpha_\ast = \frac{\mathscr{J}_1}{1-\mathscr{J}_1}$,
$\zeta_\ast = \frac{\mathscr{J}_\infty-\mathscr{J}_1}{\mathscr{J}_\infty+\mathscr{J}_1}$.
Now the large deviation rate function 
$G^{(\infty)}(\mathscr{J}_\infty, \mathscr{J}_1) 
= \mathscr{J}_\infty \lambda_\ast + \mathscr{J}_1 \mu_\ast
- F^{(\infty)}(\lambda_\ast, \mu_\ast)$
is given by
\begin{equation}\label{G2re}
\begin{split}
G^{(\infty)}(\mathscr{J}_\infty, \mathscr{J}_1) 
&=\mathscr{J}_\infty \ln\left(
\frac{\mathscr{J}_\infty-\mathscr{J}_1}{z(\mathscr{J}_\infty+\mathscr{J}_1)}\right)
+ \mathscr{J}_1  \ln\left(\frac{4(1-\mathscr{J}_1)\mathscr{J}_1^3(a+a^{-1}-2)}
{(\mathscr{J}_\infty^2-\mathscr{J}_1^2)(1-2\mathscr{J}_1)^2(z+z^{-1}-2)}\right)
\\
&- \ln\left(\frac{(1-a)(1-\mathscr{J}_1)}{1-2\mathscr{J}_1}\right).
\end{split}
\end{equation}
Since every soliton carries at least one ball, 
the ball current $\mathscr{J}_\infty$ and the soliton current $\mathscr{J}_1$ are to be considered 
in the domain $\mathscr{J}_\infty > \mathscr{J}_1$.
Similarly, from (\ref{sj}) with $l \rightarrow \infty$, 
one should suppose $\mathscr{J}_1<1$.
Note that $G^{(\infty)}(\mathscr{J}_\infty, \mathscr{J}_1) $ is {\em finite} 
at $\mathscr{J}_1= \frac{1}{2}$.

The single temperature case in the main text 
corresponds to setting $a =z$, $\mu = 0$.
In fact,  (\ref{qlim2}) enforces $\alpha_\ast = \zeta_\ast$ leading to 
$\mathscr{J}_1 = \frac{\mathscr{J}_\infty}{1+2\mathscr{J}_\infty}$
and $\mu_\ast=0$.
Then (\ref{G2re}) reduces $G^{(\infty)}(j= \mathscr{J}_\infty)$ in the main text.

\section{Microscopic definition of $\hat{\eta}^{(l)}_i(x)$}
\label{app:eta_def}

Here we outline the proof of the properties of 
the generalized current $\hat{\eta}^{(l)}_i(x)$ claimed 
in the beginning of Section \ref{ssec:generalized_currents}.
We begin by recalling the combinatorial $R$ in the crystal base theory,  
a theory of quantum groups at $q=0$,  for which 
readers are referred to  [Section 2.2, \cite{IKT12}] and the references therein. 

For a positive integer $l$, define a set $B_l =\{\alpha= (\alpha_0,\alpha_1) \in (\Z_{\ge 0})^2\mid \alpha_0+\alpha_1=l\}$.
Introduce an infinite set known as  an {\em affine crystal}
$\mathrm{Aff}(B_l) = B_l \times \zeta^{\Z} = \{\alpha \zeta^n\mid \alpha \in B_l, n \in \Z\}$,
where $\zeta$ is an indeterminate.
Define a map $R_{l,m}: \mathrm{Aff}(B_l) \otimes \mathrm{Aff}(B_m)  
\rightarrow \mathrm{Aff}(B_m) \otimes \mathrm{Aff}(B_l) $ by\footnote{Tensor product $\otimes$ in this appendix can just be 
regarded as product of sets.}
\begin{align}
R_{l,m}: & \;\alpha \zeta^a\otimes \beta \zeta^b
\mapsto \tilde{\beta}\zeta^{b+H(\alpha \otimes \beta)}\otimes \tilde{\alpha}\zeta ^{a-H(\alpha \otimes \beta)}
\label{rlm}
\end{align}
for any $a,b \in \Z$.
Here for $\alpha=(\alpha_0,\alpha_1)\in B_l$, $\beta=(\beta_0,\beta_1)\in B_m$, the image 
$\tilde{\alpha}=(\tilde{\alpha}_0,\tilde{\alpha}_1) \in B_l$,  $\tilde{\beta}=(\tilde{\beta}_0,\tilde{\beta}_1) \in B_m$
and $H(\alpha \otimes \beta) \in \Z$ are specified by 
\begin{align}
&\tilde{\alpha}_i =\alpha_i +\min(\alpha_{i+1},\beta_i)-\min(\alpha_{i},\beta_{i+1}),
\;\;
\tilde{\beta}_i = \beta_i - \min(\alpha_{i+1},\beta_i)+\min(\alpha_{i},\beta_{i+1}),
\\
&H(\alpha \otimes \beta)  =\min(\alpha_0, \beta_1)
\label{eq:Hmin}
\end{align}
with all the indices in $\Z_2$.
The map $R_{l,m}$ is called an (affine) {\em combinatorial R}.  
It is the quantum $R$ matrix for $(\text{spin $l/2$ rep} )\otimes (\text{spin $m/2$ rep})$ of
$U_q(\widehat{sl}_2)$ 
at $q=0$ with respect to the crystal base, which retains all the combinatorial essence.
The indeterminate
$\zeta$ is a remnant of the spectral parameter of the $R$ matrices.
The simpler version forgetting it (the one formally corresponding to $\zeta=1$)
is called (classical) combinatorial $R$.
In what follows the both versions will simply be denoted by $R$.
The relation (\ref{rlm}) is customarily depicted as

\begin{picture}(80,80)(-140,-20)

\put(-19,18){$\alpha \zeta^a$}\put(15,45){$\beta \zeta^b$}
\put(0,20){\vector(1,0){40}}
\put(20,40){\vector(0,-1){40}}
\put(45,18){$\tilde{\alpha}\zeta^{a-H}\qquad (H = H(\alpha\otimes \beta))$.}
\put(13,-11){$\tilde{\beta}\zeta^{b+H}$}

\end{picture}

\noindent
It satisfies the inversion and the Yang-Baxter relations
\begin{align}
&R_{m,l}R_{l,m} = \mathrm{Id},
\label{inversion}
\\
&(1 \otimes R_{k,l})(R_{k,m}\otimes 1)(1\otimes R_{l,m})
= (R_{l,m} \otimes 1)(1\otimes R_{k,m})(R_{k,l}\otimes 1),
\label{aybe}
\end{align}
which are equalities of the maps
 $\mathrm{Aff}(B_l) \otimes\mathrm{Aff}(B_m)
\rightarrow 
\mathrm{Aff}(B_l) \otimes\mathrm{Aff}(B_m)$
and 
$\mathrm{Aff}(B_k) \otimes\mathrm{Aff}(B_l) \otimes\mathrm{Aff}(B_m)
\rightarrow 
\mathrm{Aff}(B_m) \otimes\mathrm{Aff}(B_l) \otimes\mathrm{Aff}(B_k)$
for any $k,l,m \in \Z_{\ge 1}$.
Note that  these relations 
include the equality of the powers of $\zeta$ for each tensor component.
For example, the inversion relation tells
\begin{align}\label{Heq}
H(\alpha \otimes \beta ) = H(\tilde{\beta} \otimes \tilde{\alpha}).
\end{align}

Now we consider BBS.
An element $\alpha\in B_l$ can be interpreted as a capacity $l$ carrier containing $\alpha_1$ balls.
When $l=1$, it may also be regarded as a local BBS state containing $\alpha_1(=0,1)$ ball. 
The BBS on the length $L$ periodic lattice is a dynamical system 
on $B_1^{\otimes L}$.
In what follows, a BBS state $s=(s_1,\ldots, s_L) \in \{0,1\}^L$
is identified with $(1-s_1,s_1) \otimes \cdots  \otimes (1-s_L,s_L) \in B^{\otimes L}_1$,
which will also be denoted by $s_1 \otimes \cdots \otimes s_L$.

The time evolution $s'=T_l(s)$ and the associated $l^{\text{th}}$ energy $E_l(s)$ 
are defined 
by the composition of the combinatorial $R$ as follows:

\begin{picture}(200,50)(-110,-5)

\put(30,33){$s_1 \quad s_2$ }\put(30,-2){$s'_1 \quad \, s'_2$}

\put(0,18){$u \zeta^0$}
\multiput(34,28)(20,0){2}{\vector(0,-1){20}}
\put(20,20){\line(1,0){48}} \multiput(70,19.5)(3,0){9}{.}

\put(-60,0){
\put(170,33){$s_L$}\put(170,-2){$s'_L$}
\put(160,20){\vector(1,0){30}}
\put(195,18){$u \zeta^{-E_l(s)}$}
\put(174,28){\vector(0,-1){20}}
}
\end{picture}

\noindent
Here the carrier $u \in B_l$ is determined uniquely from $s=s_1\otimes \cdots \otimes s_L$ 
by the periodic boundary condition, namely,  by requiring that it comes back to 
$u$ itself after penetrating $s$ provided that the ball density is not exactly $1/2$.

Let $u' \in B_i$ be another carrier for the time evolution $T_l(s) \rightarrow T_iT_l(s)$.
Set $R: u' \zeta^0 \otimes u \zeta^0 \mapsto {\tilde u} \zeta^h \otimes {\tilde u'} \zeta^{-h}$
where $h = H(u' \otimes u)$.
From the Yang-Baxter relation we have

\begin{picture}(400,175)(-100,5)
\setlength{\unitlength}{1.2mm}

\put(0,10){\line(1,0){8}}
\put(0,20){\line(1,0){8}}

\put(11,9){$\cdots$} 
\put(12,19){$\cdots$}

\put(18,10){\line(1,0){8}}
\put(18,20){\line(1,0){8}}

\put(4,23){\vector(0,-1){16}}
\put(22,23){\vector(0,-1){16}}

\put(-3,19.5){$\scriptstyle u$}
\put(-3,9.5){$\scriptstyle u'$}

\put(12,23){$\scriptstyle s$}
\put(11,14){$\scriptstyle T_{l}(s)$}
\put(10,5){$\scriptstyle T_iT_{l}(s)$}

\put(28,19){$\scriptstyle u \zeta^{-E_{l}(s)}$}
\put(27,9){$\scriptstyle u' \zeta^{-E_{i}(T_{l}(s))}$}

\put(-3,0){
\put(44,10){\vector(1,1){10}}
\put(44,20){\vector(1,-1){10}}
\put(55,9){$\scriptstyle \tilde{u} \zeta^{-E_{l}(s)+h}$}
\put(55,19){$\scriptstyle {\tilde u'} \zeta^{-E_{i}(T_{l}(s))-h}$}
}

\put(-10,14){$=$}

\put(1,0){
\put(0.2,35){\line(1,1){10}}
\put(0.2,45){\line(1,-1){10}}
\put(-3,35){$\scriptstyle u'$}
\put(-3,45){$\scriptstyle u$}
\put(12,35){$\scriptstyle {\tilde u} \zeta^{h}$}
\put(11,45){$\scriptstyle {\tilde u'} \zeta^{-h}$}
}

\put(20,35){\line(1,0){8}}
\put(20,45){\line(1,0){8}}

\put(31,34){$\cdots$} 
\put(32,44){$\cdots$}

\put(38,35){\vector(1,0){8}}
\put(38,45){\vector(1,0){8}}

\put(24,48){\vector(0,-1){16}}
\put(42,48){\vector(0,-1){16}}

\put(32,48){$\scriptstyle s$}
\put(31,39){$\scriptstyle T_{i}(s)$}
\put(30,30){$\scriptstyle T_{l}T_{i}(s)$}

\put(47,45){$\scriptstyle {\tilde u'}\zeta^{-E_{i}(s)-h}$}
\put(47,35){$\scriptstyle {\tilde u} \zeta^{-E_{l}(T_{i}(s))+h}$}

\end{picture}

Comparing the two sides we get the commutativity $T_lT_i(s) = T_iT_l(s)$ and 
the energy conservation $E_i(s) = E_i(T_l(s))$, $E_l(s) = E_l(T_i(s))$ as is well known.

Now consider the intermediate stage where the carriers have only gone through the 
first $x-1$ local states  
$s_{<x} = s_1 \otimes \cdots \otimes s_{x-1}$.
To systematize the notation we write $u, u', h$ in the above diagram as 
$u(0), u'(0), h(0)$.
Then the corresponding diagram looks as 

\begin{picture}(400,175)(-100,5)
\setlength{\unitlength}{1.2mm}

\put(0,10){\line(1,0){8}}
\put(0,20){\line(1,0){8}}

\put(11,9){$\cdots$} 
\put(12,19){$\cdots$}

\put(18,10){\line(1,0){8}}
\put(18,20){\line(1,0){8}}

\put(4,23){\vector(0,-1){16}}
\put(22,23){\vector(0,-1){16}}

\put(-11,19.5){$\scriptstyle B_l \,\ni \,  u(0)$}
\put(-11,9.5){$\scriptstyle B_i \,\ni \, u'(0)$}

\put(12,23){$\scriptstyle s_{<x}$}
\put(11,14){$\scriptstyle T_{l}(s_{<x})$}
\put(10,5){$\scriptstyle T_iT_{l}(s_{<x})$}

\put(28,19.5){$\scriptstyle u(x) \zeta^{-E_{l}(s_{<x})}$}
\put(27,9.5){$\scriptstyle u' (x)\zeta^{-E_{i}(T_{l}(s_{<x}))}$}

\put(3.5,0.5){
\put(44,10){\vector(1,1){10}}
\put(44,20){\vector(1,-1){10}}
\put(55,9){$\scriptstyle \tilde{u}(x) \zeta^{-E_{l}(s_{<x})+h(x)}$}
\put(55,19){$\scriptstyle {\tilde u'}(x) \zeta^{-E_{i}(T_{l}(s_{<x}))-h(x)}$}
}

\put(-18,14){$=$}

\put(-4,0){
\put(0.2,35){\line(1,1){10}}
\put(0.2,45){\line(1,-1){10}}
\put(-10,45){$\scriptstyle B_l \,\ni \, u(0)$}
\put(-11,35){$\scriptstyle B_i \, \ni \, u'(0)$}
\put(12,35){$\scriptstyle {\tilde u}(0) \zeta^{h(0)}$}
\put(11,45){$\scriptstyle {\tilde u'}(0) \zeta^{-h(0)}$}
}

\put(20,35){\line(1,0){8}}
\put(20,45){\line(1,0){8}}

\put(31,34){$\cdots$} 
\put(32,44){$\cdots$}

\put(38,35){\vector(1,0){8}}
\put(38,45){\vector(1,0){8}}

\put(24,48){\vector(0,-1){16}}
\put(42,48){\vector(0,-1){16}}

\put(32,48){$\scriptstyle s_{<x}$}
\put(31,39){$\scriptstyle T_{i}(s_{<x})$}
\put(30,30){$\scriptstyle T_{l}T_{i}(s_{<x})$}

\put(47,45){$\scriptstyle {\tilde u'}(x)\zeta^{-E_{i}(s_{<x})-h(0)}$}
\put(47,35){$\scriptstyle {\tilde u}(x) \zeta^{-E_{l}(T_{i}(s_{<x}))+h(0)}$}

\end{picture}

\noindent
where $h(x) = H(u'(x) \otimes u(x))$.
The carriers $u(x)$ and $u'(x)$ (resp. ${\tilde u}(x)$ and ${\tilde u'}(x)$) 
at this position 
do not yet have to return to the initial ones 
$u(0)$ and $u'(0)$ (resp. ${\tilde u}(0)$ and ${\tilde u'}(0)$).
Define the local observables
\begin{align}
\hat{\eta}^{(l)}_i(x) = H(u'(x) \otimes u(x)),\qquad 
\hat{\eta}^{(i)}_l(x) = H({\tilde u}(x) \otimes {\tilde u'}(x)).
\label{eq:etaH}
\end{align}
Then the following properties are satisfied.
\begin{align}
\text{(i)} \;\;& \hat{\eta}^{(l)}_i(x)  = \hat{\eta}^{(i)}_l(x),
\label{symeta}
\\
\text{(ii)}\; \; &  \begin{cases}
E_l(T_i(s_{<x}))- E_l(s_{<x}) +  {\hat \eta}^{(i)}_l(x) - {\hat \eta}^{(i)}_l(0)=0,
\\
E_i(T_l(s_{<x}))- E_i(s_{<x}) +{\hat \eta}^{(l)}_i(x) - {\hat \eta}^{(l)}_i(0)=0.
\end{cases}
\label{etacon}
\end{align}
In fact  (i) is a consequence of (\ref{Heq}).
As for (ii) both relations follow simultaneously by comparing 
the powers of $\zeta$ in the above diagram using (i).
The upper relation, for instance, is 
nothing but the space $[0,x]$-integrated 
equation of continuity for $E_l$ with respect to the time evolution $T_i$,
where $\hat{\eta}^{(i)}_l(x)$ plays the role of local current at $x$.

As an example, $\hat{\eta}^{(3)}_2(x)$ 
associated with $T_2T_3$ are given in red letters in the
top panel of Fig.~\ref{fig:eta}.
On the other hand, $\hat{\eta}^{(2)}_3(x)$ associated with $T_3T_2$ are given similarly in the bottom panel of Fig.~\ref{fig:eta}.
One can observe the equality $\hat{\eta}^{(3)}_2(x) = \hat{\eta}^{(2)}_3(x)$ everywhere.
In this example, $\hat{\eta}^{(2)}_3(x) = \hat{\eta}^{(2)}_\infty(x)$ holds and it coincides with 
the carriers for $T_2$, which is indeed the ball current. 

\section*{Bibliography}
\bibliography{correl}
\end{document}